\documentclass[preprint,amsmath,amsfonts,amssymb,aps,prx,preprintnumbers,superscriptaddress]{revtex4-2}

\usepackage[utf8]{inputenc}
\usepackage[T1]{fontenc}
\usepackage{lmodern}
\usepackage{graphicx}
\usepackage{dcolumn}
\usepackage{bm}
\usepackage{textcomp}
\usepackage{ulem}
\usepackage{ifpdf}

\usepackage{color}
\definecolor{red}{rgb}{1,0,0}
\definecolor{blue}{rgb}{0,0,1}
\definecolor{darkred}{rgb}{0.6,0,0}
\definecolor{darkblue}{rgb}{0,0,.6}
\definecolor{darkgreen}{rgb}{0,0.5,0}

\ifpdf
  \usepackage{epstopdf}
  \usepackage[pdftex,unicode,pdfstartview={FitH},pdfborder={0 0 0}]{hyperref}
  \usepackage{hypcap}
\else
  \usepackage[hypertex]{hyperref}
\fi
\hypersetup{
  bookmarksnumbered = true,
  colorlinks = true, linkcolor = darkblue,
  citecolor = darkblue, filecolor = darkblue,
  menucolor = darkblue, urlcolor = darkblue
}

\usepackage{etoolbox}
\makeatletter
\patchcmd{\frontmatter@RRAP@format}{(}{}{}{}
\patchcmd{\frontmatter@RRAP@format}{)}{}{}{}
\renewcommand\Dated@name{ *A.S.B. email: anvar.baimuratov@lmu.de \\ ***A.H. email: alexander.hoegele@lmu.de}
\makeatother

\newcolumntype{C}{>{$\displaystyle}c<{$}}
\newcolumntype{R}{>{$\displaystyle}r<{$}}

\begin{document}

\title{Moir\'e excitons in \texorpdfstring{MoSe$_2$-WSe$_2$}{MoSe2-WSe2} heterobilayers and heterotrilayers}

\author{Michael F{\"o}rg}
\affiliation{Fakult\"at f\"ur Physik, Munich Quantum Center, and
  Center for NanoScience (CeNS), Ludwig-Maximilians-Universit\"at
  M\"unchen, Geschwister-Scholl-Platz 1, 80539 M\"unchen, Germany}

\author{Anvar~S.~Baimuratov*}
\affiliation{Fakult\"at f\"ur Physik, Munich Quantum Center, and
  Center for NanoScience (CeNS), Ludwig-Maximilians-Universit\"at
  M\"unchen, Geschwister-Scholl-Platz 1, 80539 M\"unchen, Germany}

\author{Stanislav~Yu.~Kruchinin}
\affiliation{Center for Computational Materials Sciences, Faculty of Physics, University of Vienna, Sensengasse 8/12, 1090 Vienna, Austria}
\affiliation{Nuance Communications Austria GmbH, Technologiestraße 8, 1120 Wien}

\author{Ilia~A.~Vovk}
\affiliation{Center of Information Optical Technology, ITMO University, Saint Petersburg 197101, Russia}

\author{Johannes Scherzer}
\affiliation{Fakult\"at f\"ur Physik, Munich Quantum Center, and
  Center for NanoScience (CeNS), Ludwig-Maximilians-Universit\"at
  M\"unchen, Geschwister-Scholl-Platz 1, 80539 M\"unchen, Germany}

\author{Jonathan F{\"o}rste}
\affiliation{Fakult\"at f\"ur Physik, Munich Quantum Center, and
  Center for NanoScience (CeNS), Ludwig-Maximilians-Universit\"at
  M\"unchen, Geschwister-Scholl-Platz 1, 80539 M\"unchen, Germany}

\author{Victor Funk}
\affiliation{Fakult\"at f\"ur Physik, Munich Quantum Center, and
  Center for NanoScience (CeNS), Ludwig-Maximilians-Universit\"at
  M\"unchen, Geschwister-Scholl-Platz 1, 80539 M\"unchen, Germany}

\author{Kenji Watanabe}
\affiliation{Research Center for Functional Materials, National Institute for Materials Science, 1-1 Namiki, Tsukuba 305-0044, Japan}

\author{Takashi Taniguchi}
\affiliation{International Center for Materials Nanoarchitectonics, 
National Institute for Materials Science, 1-1 Namiki, Tsukuba 305-0044, Japan}

\author{Alexander H{\"o}gele*}
\affiliation{Fakult\"at f\"ur Physik, Munich Quantum Center, and
  Center for NanoScience (CeNS), Ludwig-Maximilians-Universit\"at
  M\"unchen, Geschwister-Scholl-Platz 1, 80539 M\"unchen, Germany}
\affiliation{Munich Center for Quantum Science and Technology (MCQST),
  Schellingtra\ss{}e 4, 80799 M\"unchen, Germany}

\date{}

\begin{abstract}
Layered two-dimensional materials exhibit rich transport and optical phenomena in twisted or lattice-incommensurate heterostructures with spatial variations of interlayer hybridization arising from moir\'e interference effects. Here, we report experimental and theoretical studies of excitons in twisted heterobilayers and heterotrilayers of transition metal dichalcogenides. Using MoSe$_2$-WSe$_2$ stacks as representative realizations of twisted van der Waals bilayer and trilayer heterostructures, we observe contrasting optical signatures and interpret them in the theoretical framework of interlayer moir\'e excitons in different spin and valley configurations. We conclude that the photoluminescence of MoSe$_2$-WSe$_2$ heterobilayer is consistent with joint contributions from radiatively decaying valley-direct interlayer excitons and phonon-assisted emission from momentum-indirect reservoirs that reside in spatially distinct regions of moir\'e supercells, whereas the heterotrilayer emission is entirely due to momentum-dark interlayer excitons of hybrid-layer valleys. Our results highlight the profound role of interlayer hybridization for transition metal dichalcogenide heterostacks and other realizations of multi-layered semiconductor van der Waals heterostructures.
\end{abstract}

\maketitle

\section{Introduction}
Heterostructures of layered two-dimensional materials exhibit rich transport and optical phenomena. In twisted or lattice-incommensurate heterobilayers (HBLs), laterally modulated van der Waals interactions give rise to spatial variations in the degree of interlayer hybridization on the characteristic length scale of the moir\'e interference pattern \cite{Wu2017,WangYu2017,Yu2017,Wu2018,Yu2018,Ruiz2019}. The formation of moir\'e superlattices has profound effects on the electronic band structure, as evidenced by the emergence of correlated transport phenomena in flat bands of twisted bilayer \cite{Cao2018-1,Cao2018-2} and trilayer \cite{Chen2019-1,Chen2019-2} graphene, or detected optically in twisted homobilayers \cite{Shimazaki2020} and aligned HBLs \cite{Tang2020,Regan2020} of transition metal dichalcogenides (TMDs). The latter also exhibit rich moir\'e signatures in the optical spectra of intralayer \cite{Plochocka2018} and interlayer \cite{seyler2019,tran2019,jin2019,alexeev2019} excitons formed by Coulomb attraction among layer-locked and layer-separated electrons and holes.

In MoSe$_2$-WSe$_2$ HBL, a prominent representative of TMD heterostacks, the interlayer exciton photoluminescence (PL) is observed well below the intralayer features of monolayer MoSe$_2$ and WSe$_2$ constituents \cite{Rivera2015}. The PL energy is consistent with a staggered band alignment \cite{Kang2013} which separates electrons and holes into the conduction and valence bands of MoSe$_2$ and WSe$_2$, respectively. In accord with layer separation, interlayer excitons exhibit strongly prolonged radiative lifetimes \cite{Rivera2015} and reduced oscillator strength \cite{Forg2019}. Despite numerous experimental and theoretical studies of MoSe$_2$-WSe$_2$ HBLs, the origin of the lowest energy PL remains a subject of debate \cite{Deilmann2020}. While the majority of experimental studies interpret the HBL emission in terms of zero-momentum interlayer excitons with $K$ or $K'$ valley electrons and holes in MoSe$_2$ and WSe$_2$ \cite{Rivera2015,Hsu2018,Ciarrocchi2019,seyler2019,tran2019,Forg2019,ZhangTutuc2019,Wang2020giant,Calman2020,Joe2019,Delhomme2020}, others invoke excitons built from hybridized HBL conduction band states at $Q$ pockets \cite{Nayak2017,Miller2017,Hanbicki2018}, located roughly halfway between the center of the first Brillouin zone at $\Gamma$ and $K$ or $K'$ valleys. Band structure calculations indeed suggest that hybridization of states near $Q$ conduction band and $\Gamma$ valence band of MoSe$_2$ and WSe$_2$ gives rise to strong energy renormalization upon HBL formation \cite{WangYu2017,Gillen2018,Torun2018} which might turn either $QK$ or $Q\Gamma$ interlayer excitons, composed of electrons at $Q$ and holes at $K$ or $\Gamma$, into the lowest energy states. 

Additional complication arises in the presence of moir\'e effects. In moir\'e-modulated HBLs, electronic states exhibit valley-contrasting energy shifts upon interlayer hybridization, with states in $K$ and $K'$ valleys being less susceptible to energy reducing interactions than the conduction band states at $Q$ or the valence band states at $\Gamma$. This effect, analogous to the origin of the direct-to-indirect band gap cross-over in TMD monolayers and bilayers \cite{Splendiani2010,Mak2010,Liu2015a}, should also impact the band structure of HBLs \cite{WangYu2017} yet has been mostly neglected in the context of moir\'e excitons \cite{Wu2017,Yu2017,Wu2018,Yu2018,Ruiz2019}. Interlayer hybridization is expected to play an even more prominent role in heterotrilayer (HTL) systems with native homobilayers. For the explicit case of MoSe$_2$-WSe$_2$ HTLs, one would expect sizable hybridization effects between the MoSe$_2$ bilayer band edge states at $Q$ and their counterparts in monolayer WSe$_2$, rendering the overall heterostructure an indirect band gap semiconductor.

\section{Results}

\subsection{MoSe2-WSe2 heterobilayer and heterotrilayer in cryogenic spectroscopy}
\vspace{-12pt}

Motivated by the contrasting behavior anticipated for momentum direct and indirect band edge interlayer excitons in MoSe$_2$-WSe$_2$ HBL and HTL, we performed optical spectroscopy studies of the corresponding moir\'e heterostructures on the same sample.  The heterostack was connected to a charge-reservoir for voltage control of capacitive doping. To this end, a MoSe$_2$ crystal with monolayer and bilayer regions was stacked onto a WSe$_2$ monolayer by dry viscoelastic stamping \cite{hot-pick-up},  encapsulated from both sides by hexagonal boron nitride (hBN) and stamped into contact with a gold electrode with gate voltage referenced against a grounded layer of silver capped by SiO$_2$ (Supplementary Note~1). The MoSe$_2$ crystal with a native bilayer region in 2H or AA' stacking was twisted away from parallel R-type alignment by about $4^\circ$ with respect to the WSe$_2$ monolayer. At such relatively large angles, we expect the moir\'e heterostructure to be robust against reconstruction \cite{Carr2018,Enaldiev2019,Holler2020} and thus to contrast previous studies of MoSe$_2$-WSe$_2$ HBLs carefully aligned for zero twist angle in R-type stacking \cite{Ciarrocchi2019,Joe2019} as well as moir\'e-free HBLs obtained from chemical vapor deposition with lattice-mismatch relaxation and inherent alignment \cite{Hsu2018,Forg2019}. 

\begin{figure}[t!]
\vspace{-18pt}
\includegraphics[scale=1.04]{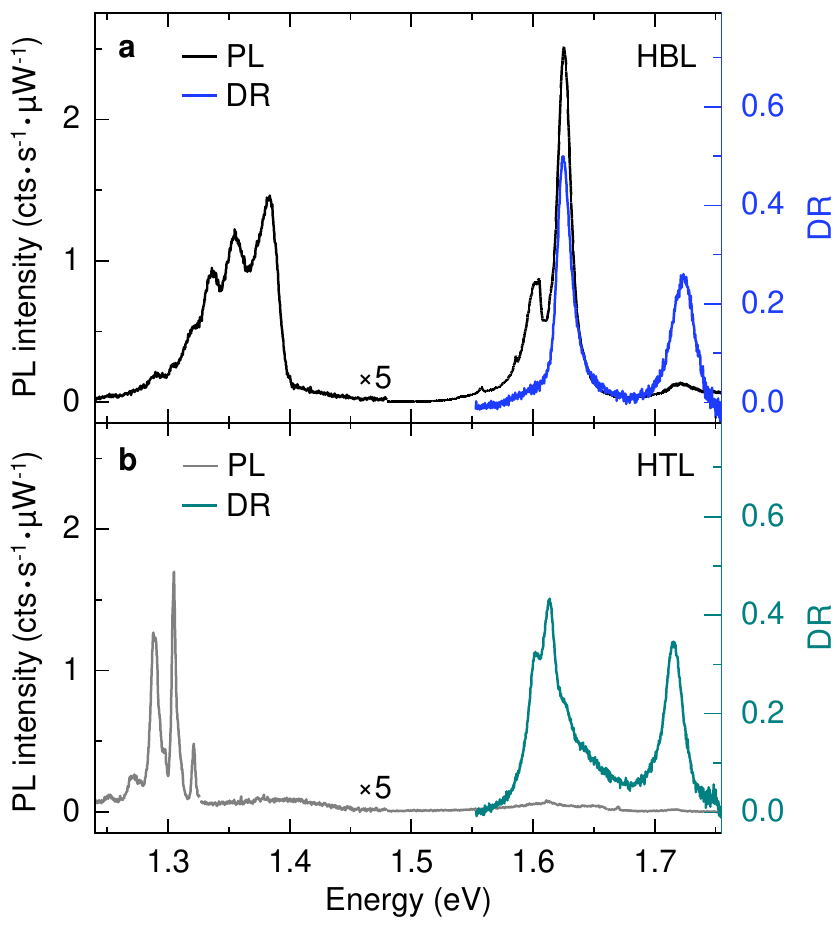}
\vspace{-10pt}
\caption{\textbf{Cryogenic photoluminescence and differential reflectivity spectra of MoSe$_2$-WSe$_2$ heterobilayer and heterotrilayer.} \textbf{a} and \textbf{b}, Photoluminescence (black and grey) and differential reflectivity (blue and dark cyan) spectra of twisted HBL and HTL MoSe$_2$-WSe$_2$ at $3.2$~K. The luminescence was excited with linearly polarized excitation laser at $1.85$~eV and scaled in intensity below $1.47$~eV by a factor of 5 in both graphs.} \label{fig1}
\end{figure}

Cryogenic PL and differential reflectivity (DR) spectra of the HBL and HTL regions at $3.2$~K and zero gate voltage are shown in Fig.~\ref{fig1}a and b, respectively. The DR features in the spectral range between $1.55$~eV and $1.75$~eV are consistent with absorption characteristics of neutral intralayer excitons, and the vanishingly small trion feature in Fig.~\ref{fig1}a indicates operation close to charge-neutrality (Supplementary Note~1). Whereas the two dominant DR peaks of the HBL spectrum in Fig.~\ref{fig1}a essentially reflect the respective MoSe$_2$ and WSe$_2$ monolayer transitions around $1.6$ and $1.7$~eV, the HTL spectrum in Fig.~\ref{fig1}b is different. Compared to the HBL spectrum, it exhibits a red-shift of the WSe$_2$ intralayer exciton peak by $8$~meV because of Coulomb screening by the additional MoSe$_2$ layer, and a rich structure around the MoSe$_2$ absorption peak with possible contributions from interlayer excitons of bilayer MoSe$_2$ \cite{Horng2018} as well as moir\'e miniband effects \cite{Ruiz2019} in the twisted HTL.

Within the same energy range, the cryogenic PL is consistently dominated by intralayer excitons. Remarkably, the intralayer MoSe$_2$ and WSe$_2$ peaks in the HBL spectrum of Fig.~\ref{fig1}a are nearly completely quenched in the HTL spectrum of Fig.~\ref{fig1}b, indicating for the latter drastically suppressed hot luminescence due to enhanced population relaxation into lowest-energy interlayer exciton levels. This observation is in accord with the theoretical prediction of increased charge transfer efficiency via hybridized $Q$ and $\Gamma$ states in heterostructures \cite{WangYu2017}. 

Another striking difference in the PL of the heterostacks is evident in the spectra of Fig.~\ref{fig1}a and b for interlayer excitons, with PL emission below $1.40$ and $1.33$~eV from HBL and HTL regions, respectively. The PL characteristics depend on the heterostack position, as confirmed by lateral displacement of the sample with respect to fixed confocal excitation and detection spots. Upon transition from the HBL to the HTL region, the set of the HTL peaks below $1.35$~eV emerges at the expense of the higher energy HBL peaks with emission energy above $1.35$~eV (Supplementary Note~2). At each heterostack site, the overall multi-peak PL structure of HBL and HTL is mostly preserved upon the variation in the gate voltage (Supplementary Note~1) and excitation power down to $300$~nW (Supplementary Note~3). Consistent with finite twist angle, the multi-peak PL of the HBL below $1.40$~eV, with a peak separation of $30$~meV between the two highest energy peaks and $15$~meV between other consecutive peaks (Supplementary Note~3), is reminiscent of rich MoSe$_2$-WSe$_2$ moir\'e spectral features \cite{tran2019} rather than of simple spectra from aligned HBLs \cite{Ciarrocchi2019,ZhangTutuc2019,Wang2020giant,Calman2020,Joe2019,Delhomme2020}. Remarkably, the HTL PL, with a similar peak spacing of $15$~meV, is strikingly similar to the cryogenic PL from native bilayer WSe$_2$ \cite{Lindlau2017BL} (Supplementary Note~4).

In time-resolved PL, HBL and HTL PL exhibit similar decay dynamics (Supplementary Note~5). Spectral sampling of the PL decay characteristics indicates the presence of at least three decay channels without conclusive dependence on the emission energy for both HBL and HTL emission. Throughout the spectral band of interlayer excitons, the PL decay exhibits three decay timescales of $3$, $12$ and $480$~ns for HBL and $1$, $12$ and $300$~ns for HTL. For both heterostacks, PL decay was dominated by the slow component (with a weight of $89\%$ and $80\%$ in HBL and HTL, respectively) with contributions of the intermediate (fast) decay channel of $8\%$ and $13\%$ ($3\%$ and $7\%$) to the HBL and HTL emission, respectively. These time scales are in accord with previous studies of MoSe$_2$-WSe$_2$ HBL \cite{Miller2017,Jiang2018,Forg2019,Choi2020} and subject to different and partly competing interpretations.

\subsection{Theory of excitons in R-stacked MoSe2-WSe2 heterobilayer and heterotrilayer}
\vspace{-12pt}

The differences in the PL spectra of Fig.~\ref{fig1}a and b suggest different origins for the interlayer exciton PL in MoSe$_2$-WSe$_2$ HBL and HTL. To provide a basis for the interpretation of our observations, we performed numerical calculations of the band structure and exciton $g$-factors with density functional theory (DFT) in generalized gradient approximation (Supplementary Notes~6 and 7). Assuming that the twist angle is sufficiently small to employ local band structure approximation \cite{Yu2015,Wu2018}, we restrict our analysis to three high symmetry points of the moir\'e superlattice in each heterostructure with stackings indicated in Fig.~\ref{fig2}a (Supplementary Note~6). Using the band structure results from DFT, we employed the Wannier exciton model in the effective mass approximation \cite{Berghaeuser2014} to calculate the energies of intralayer and interlayer excitons in different spin-valley configurations.

In the top panels of Fig.~\ref{fig2}b and c we show the oscillator strength, essentially determined by the squared modulus of the coordinate operator matrix elements, for direct $KK$ exciton transitions in different R-type stackings of HBL and HTL. For all stackings, interlayer excitons exhibit at least two orders of magnitude lower oscillator strengths than their intralayer counterparts \cite{Gillen2018} with dipolar selection rules in agreement with the group theory analysis of R-type HBL \cite{Yu2018,Forg2019}. In accord with previous calculations for HBLs, we find the lowest-energy $KK$ interlayer exciton for A'B' \cite{Wu2018,Gillen2018} and energetically higher excitons for AA and AB' stackings. In all HBL stackings of R-type registry considered here, the lowest $KK$ interlayer exciton is spin-like (i.e. in collinear electron spin orientation of occupied conduction band and unoccupied valence band states, and equivalent to spin-singlet configuration in the electron-hole notation), about $20$~meV below its spin-unlike counterpart (or spin-triplet electron-hole pair). In AA stacking, the spin-like state has the largest oscillator strength, whereas for spin-unlike states only the $KK$ exciton in A'B' stacking has a sizable oscillator strength in agreement with previous DFT results \cite{Gillen2018,Torun2018}. 

For the HTL, our calculations predict an increase in the number of conduction bands associated with lowest energy excitons due to the additional MoSe$_2$ layer. As such, interlayer $KK$ excitons can be grouped according to the localization of the conduction band electron in one of the MoSe$_2$ layers. For electrons localized on the MoSe$_2$ layer with immediate proximity to WSe$_2$ (full symbols in Fig.~\ref{fig2}c), the corresponding interlayer excitons feature similar energies (with a small red-shift due to modified screening) and oscillator strengths as in the HBL system. Additional interlayer states arise from excitons with the electron localized in the upper MoSe$_2$ layer (open symbols in Fig.~\ref{fig2}c). Their energetic ordering, with spin-unlike configuration again being lowest, and dipolar selection are identical to $KK$ interlayer excitons in HBLs of H-type registry \cite{Yu2018,Forg2019}. However, the corresponding transitions have a drastically inhibited oscillator strengths due to a reduced wavefunction overlap between the electron and hole in the topmost MoSe$_2$ and the bottom WSe$_2$ layer and thus should not contribute sizeably to the PL of HTL \cite{BrotonsGisbert2019}. Based on our analysis, we rule out excitons composed from electrons and holes that are locked in distant layers as candidates for bright PL emission in the red-most part of the HTL spectrum.

In addition to $KK$ excitons, our calculations yield the energies of momentum-indirect $QK$, $Q\Gamma$ and $K\Gamma$ excitons (bottom panels of Fig.~\ref{fig2}b and c) composed from electrons in $Q$ (or $Q'$) and $K$ as well as holes at $K$ or $\Gamma$. Note that the notion of oscillator strength is meaningless for momentum-indirect excitons without direct radiative decay pathways. The energetic ordering of interlayer excitons with zero and finite center-of-mass momentum differs substantially in HBL and HTL systems: whereas our calculations predict energetic proximity for $KK$, $QK$ and $K\Gamma$ states in HBLs, finite-momentum $QK$ and $Q\Gamma$ states in HTL are energetically well below the direct $KK$ states, with an energy difference in the order of $200$~meV. This trend is well known for monolayer and bilayer TMDs, where the states at $K$ are much less sensitive to the addition of one layer than the states at $Q$ and $\Gamma$ \cite{Splendiani2010,Mak2010,Liu2015a,Deilmann2019}. For the HTL, strong interlayer hybridization should result in efficient layer coupling as opposed to layer locking \cite{BrotonsGisbert2019}. The respective experimental signature of enhanced relaxation from intralayer to interlayer exciton states is the strong suppression of the HTL PL around the MoSe$_2$ intralayer resonance at $1.62$~eV in Fig.~\ref{fig1}b.

\vspace{12pt}
\subsection{Power-dependent photoluminescence and degree of circular polarization of MoSe2-WSe2 heterobilayer and heterotrilayer}
\vspace{-12pt}

We find experimental support for our theoretical description of HBL and HTL excitons by probing the PL and the degree of circular polarization ($\mathrm{P_C}$) as a function of excitation power. The corresponding results are shown in Fig.~\ref{fig3}a, b and c, d for HBL and HTL, respectively. Upon increasing excitation power from $0.1$ to $100~\mu$W, the HBL spectrum develops a pronounced shoulder above $1.40$~eV with vanishing $\mathrm{P_C}$ (indicated by the dashed black line in Fig.~\ref{fig3}a and b). This feature is consistent with hot luminescence from energetically higher states with $z$-polarized in-plane emission collected by our objective with high numerical aperture and corresponding collection solid angle. Our theory provides spin-unlike and spin-like interlayer excitons in AA and AB' stacking, respectively, as two potential reservoirs for this emission ($z$-polarized states in Fig.~\ref{fig2}b).

In contrast, the brightest HBL peaks between $1.32$ and $1.40$~eV with a positive degree of circular polarization are present down to lowest excitation powers. They are consistent with spin-like (spin-unlike) $KK$ interlayer excitons in AA (A'B') stacking (states in Fig.~\ref{fig2}b with $\sigma^+$ polarization). As we observe no sign reversal in $\mathrm{P_C}$ as expected for the lowest-energy spin-like $KK$ interlayer exciton in A'B' stacking (state in Fig.~\ref{fig2}b with $\sigma^-$ polarization) and reported previously for structured HBL PL \cite{tran2019}, strong contribution from A'B' stacking is unlikely in our sample. This implies that AA domains dominate the HBL PL in rigid moir\'e supercells, although A'B' regions should be at least of comparable size \cite{Carr2018}. In the presence of reconstruction, one would expect predominance of energetically favored A'B' and AB' triangular domains of comparable area \cite{Carr2018,Enaldiev2019,Weston2019}. Without reconstruction, on the other hand, a reversal in the energetic ordering of AA and A'B' interlayer excitons at finite twist angles, as predicted recently by theory for R-type MoSe$_2$-WSe$_2$ heterostructures \cite{Enaldiev2020}, would satisfactorily explain the observation.

According to our theory analysis, the structure of HTL PL is of different origin. The HTL data in Fig.~\ref{fig3}c and d reveal $KK$ interlayer exciton states through power-activated hot luminescence at the higher-energy side of the spectrum with finite circular degrees of polarization (indicated by the dashed red lines in Fig.~\ref{fig3}c and d). The respective emission peaks at $1.35$ and $1.38$~eV can be ascribed to $KK$ interlayer excitons with $\sigma^+$ polarization in (A'B')A' or (AA)A' stackings (red-colored states in Fig.~\ref{fig2}c). Similar to the HBL, the HTL spectrum exhibits upon $100~\mu$W excitation power an additional higher-energy hot luminescence peak at $1.43$~eV with vanishing $\mathrm{P_C}$ (black dashed line in Fig.~\ref{fig3}c and d). The respective candidate from theory is the $KK$ reservoir in (AB')A' stacking with $z$-polarized transition, as the (AA)A' interlayer excitons with the same selection rules are dismissed due to vanishingly small oscillator strength (black-colored states in Fig.~\ref{fig2}c). On the low energy side, the peaks around $1.30$~eV below the energy of $KK$ hot-luminescence can result as phonon sidebands from momentum-dark $QK$ or $Q\Gamma$ interlayer exciton reservoirs (states in the bottom panel of Fig.~\ref{fig2}c shown in green and orange). Without removing the ambiguity in the assignment of the lowest-energy reservoir to $QK$ or $Q\Gamma$, this scenario explains the similarity in the spectral shape of HTL PL and the PL of native bilayer WSe$_2$ originating from momentum-indirect excitons \cite{Lindlau2017BL,Foerste2020}.

\vspace{12pt}
\subsection{Magneto-luminescence of MoSe2-WSe2 heterobilayer and heterotrilayer}
\vspace{-12pt}

Additional insight into the origin of HBL and HTL peaks is obtained from magneto-luminescence experiments in Faraday configuration and theoretical analysis. The experimental dispersion of the PL peaks in external magnetic field applied perpendicular to the heterostructure is shown in Fig.~\ref{fig4}. The solid black lines indicate linear energy shifts recorded for $\sigma^+$ and $\sigma^-$ circularly polarized PL as a function of magnetic field. From this set of data, we determine the respective $g$-factors using the relation $\Delta E= g \mu_B B$, where $\Delta E$ is the energy splitting between $\sigma^+$ and $\sigma^-$ polarized peaks proportional to the interlayer exciton $g$-factor, $\mu_B$ is the Bohr magneton, and $B$ is the magnetic field. For the HBL, the extracted $g$-factors range between $-4.2$ and $-6.2$ with the same sign as for WSe$_2$ intralayer excitons, whereas the HTL peaks exhibit $g$-factors between $-12$ and $-13$. In combination with observations described above and ab-initio calculations of $g$-factors for various spin-valley configurations of interlayer excitons in high-symmetry HBL and HTL stackings (Supplementary Note 7), these values suggest the following picture for MoSe$_2$-WSe$_2$ HBL and HTL stacks twisted away from ideal R-type registry.

\vspace{+15pt}
\section{Discussion}

First, we note that the experimental $g$-factors determined for the HBL peaks from the data of Fig.~\ref{fig4}a and b ($-6.2 \pm 0.8$, $-4.2 \pm 0.8$ and $-5.5 \pm 0.8$) are consistent with previous studies of aligned MoSe$_2$-WSe$_2$ heterostructures in R-type registry with absolute values in the range from $6.1$ to $8.5$ \cite{Ciarrocchi2019,seyler2019,Joe2019}. They clearly contrast the interlayer exciton $g$-factor values between $15$ and $16$ in HBLs of H-type registry \cite{nagler2017giant,seyler2019,Wang2020giant,Delhomme2020}. For the respective $KK$ interlayer excitons in R-type HBLs, our calculations (Supplementary Note~7) predict an absolute $g$-factor value close to $6$ and opposite signs for the degrees of circular polarization in AA and A'B' stackings (with negative and positive $\mathrm{P_C}$, respectively), in agreement with a similar theoretical analysis of interlayer exciton $g$-factor values and signs \cite{Wozniak2020,Xuan2020}. 

We proceed by discussing possible origins of the structured HBL emission. In the framework of moir\'e excitons, multi-peak PL has been ascribed to interlayer exciton states confined in moir\'e quantum wells \cite{tran2019,Choi2020}. Assuming that finite twist indeed reverses the energetic ordering of interlayer excitons in A'B' and AA stackings \cite{Enaldiev2020}, the pronounced peaks of HBL PL below $1.40$~eV with positive $\mathrm{P_C}$ and negative $g$-factors in the order of $-6$ would correspond to quantum well states of zero-momentum $KK$ moir\'e excitons in AA stacked regions. This interpretation, however, is only plausible if the redistribution of interlayer exciton population among moir\'e quantum well sub-bands is bottlenecked. In the presence of population relaxation and thermal redistribution on timescales of the order or below $100 - 300$~ns (longest decay components in PL), one would expect the PL from quantum well excited states to disappear or at least to diminish in intensity due to insufficient exciton population at low excitation powers and low temperature. Instead, we always find the highest energy peak to exhibit the highest intensity down to lowest excitation powers. 
An alternative interpretation invokes $KK$ interlayer excitons trapped in disorder potentials. Within this scenario, the highest energy peak would stem from $KK$ excitons in AA stacked regions decaying via their respective $\sigma^+$ polarized radiative channel on the timescale of a few nanoseconds, whereas lower energy peaks with similar $g$-factors and $\mathrm{P_C}$ would reflect the respective defect-bound states with prolonged lifetimes. This scenario, however, is in conflict with the observation of spectrally independent PL decay, dismissing defect-trapped interlayer excitons with reduced energy as the primary source of the structured PL. The scenario of an energetically homogeneous distribution of localization by disorder over the entire spectral emission window seems even less plausible.

\begin{figure}[t!]
\vspace{-18pt}
\includegraphics[scale=0.98]{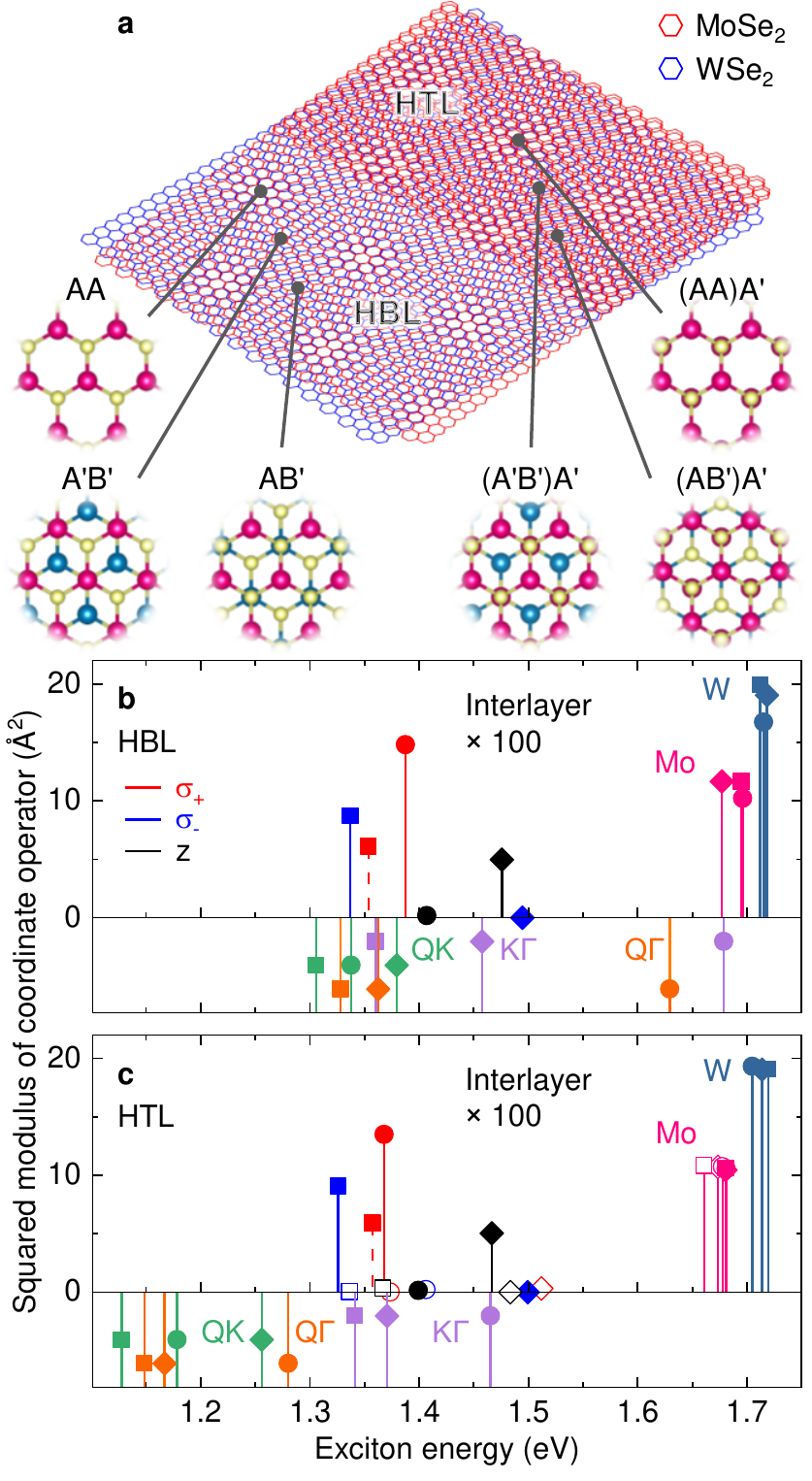}
\vspace{-15pt}
\caption{\textbf{Theory of excitons in high-symmetry stackings of MoSe$_2$-WSe$_2$ heterobilayer and heterotrilayer.} \textbf{a}, Schematics of twisted HBL and HTL MoSe$_2$-WSe$_2$ with different high-symmetry stackings. \textbf{b} and \textbf{c}, Energy and squared sum of coordinate operator (multiplied by 100 below $1.55$~eV) for intralayer and interlayer excitons in HBL and HTL, respectively, calculated for three different stackings. Filled squares, circles, and diamonds denote A'B', AA, and AB', and (A'B')A', (AA)A', and (AB')A' stackings in HBL and HTL, respectively. Empty symbols indicate corresponding HTL excitons with electrons residing in the top-most MoSe$_2$ layer. For zero-momentum $KK$ interlayer excitons (top panels) we indicate the spin configuration by solid and dashed lines for spin-like and spin-unlike states (corresponding to spin-singlet and spin-triplet excitons), and the polarization of the respective exciton emission by red ($\sigma^+$), blue ($\sigma^-$) and black (in-plane $z$) colors. The bottom panels show the energy of finite-momentum interlayer excitons in $QK$ (green), $Q\Gamma$ (orange) and $K\Gamma$ (violet) configurations without direct radiative transitions.}
\label{fig2}
\end{figure}

\begin{figure}[t]
\vspace{-18pt}
\includegraphics[scale=1.04]{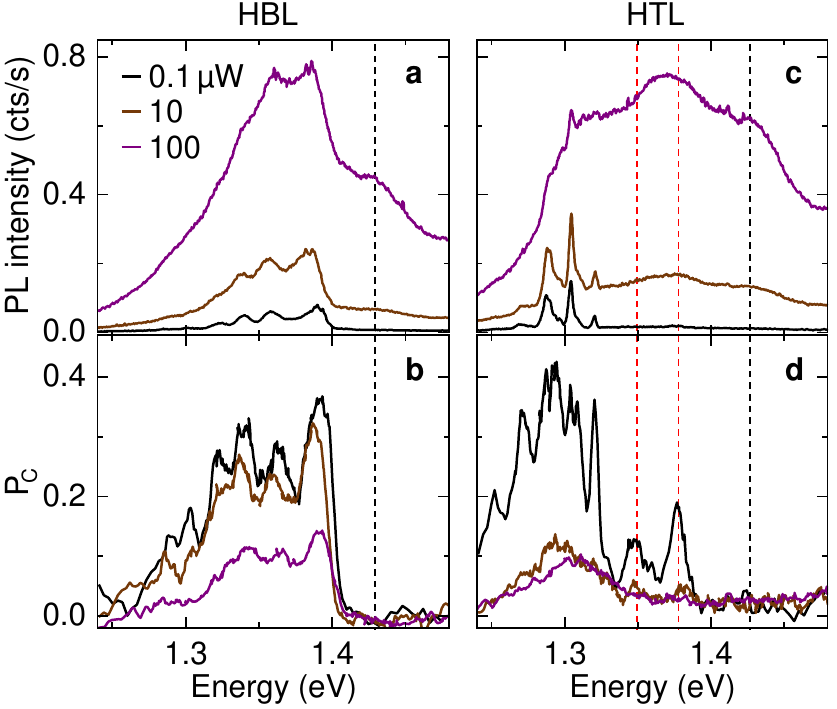}
\vspace{-10pt}
\caption{\textbf{Power-dependent photoluminescence and degree of circular polarization of MoSe$_2$-WSe$_2$ heterobilayer and heterotrilayer.} \textbf{a} and \textbf{b}, Photoluminescence spectra and degrees of circular polarization ($\mathrm{P_C}$) for twisted HBL MoSe$2$-WSe$2$ at $0.1$ (black), $10$ (brown) and $100~\mu$W (purple) excitation power. \textbf{c} and \textbf{d}, Same for HTL MoSe$2$-WSe$2$. Dashed lines indicate hot luminescence at the higher energy side of the interlayer exciton spectrum with zero (black) and finite (red) $\mathrm{P_C}$.}
\label{fig3}
\end{figure}

\begin{figure}[t!]
\vspace{-18pt}
\includegraphics[scale=1.04]{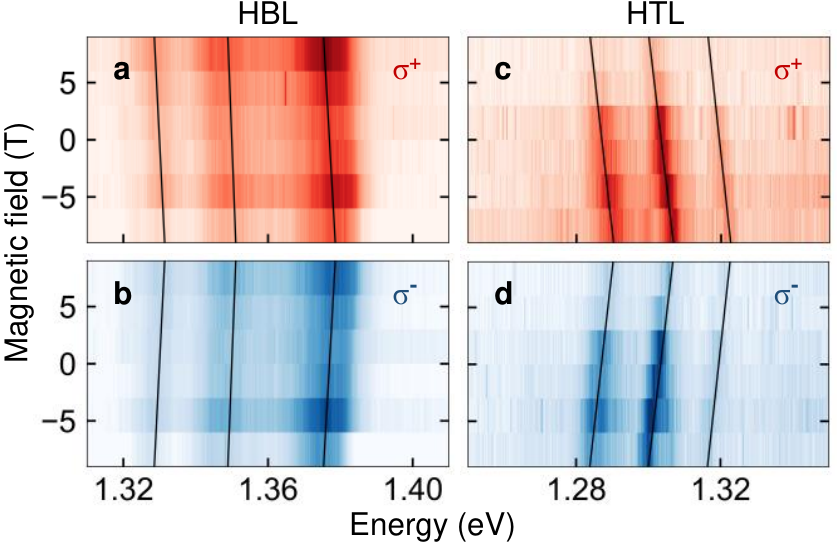}
\vspace{-10pt}
\caption{\textbf{Valley Zeeman shift of excitons in MoSe$_2$-WSe$_2$ heterobilayer and heterotrilayer.} \textbf{a} and \textbf{b}, Magneto-luminescence of twisted HBL MoSe$2$-WSe$2$ for linear excitation and $\sigma^+$ (red) and $\sigma^-$ (blue) circularly polarized detection, respectively. \textbf{c} and \textbf{d}, Same for HTL MoSe$2$-WSe$2$. The solid lines show magneto-induced energy shifts of HBL and HTL peaks with $g$-factors (from high to low energy) and error bars from least-square linear fits of $-6.2 \pm 0.8$, $-4.2 \pm 0.8$ and $-5.5 \pm 0.8$, and $-12.0 \pm 2.0$, $-12.0 \pm 0.8$ and $-13.0 \pm 0.8$, respectively.}
\label{fig4}
\end{figure}

Finally, the multi-peak structure of HBL PL can be attributed to the joint emission from zero-momentum and finite-momentum interlayer exciton reservoirs \cite{Miller2017,Forg2019}. In addition to spin-like $KK$ excitons with a $g$-factor of $6$, our theory identifies spin-like $K\Gamma$ and spin-unlike $Q'\Gamma$ with respective $g$-factors of $4$ and $5$ as candidates for the lower-energy HBL peaks. Note that theory finds these states in close energetic proximity to spin- and momentum-bright $KK$ interlayer excitons (Fig.~\ref{fig2}b). In this framework, the corresponding HBL peaks would qualify as phonon sidebands of $K\Gamma$ or $Q'\Gamma$ (or both), and the peak spacings of $30$ and $15$~meV would reflect the energies of optical and acoustic phonons, or higher-order combinations thereof \cite{Lindlau2017BL,Forg2019}. This, however, holds only for staggered A'B' and AB' stackings that favor energy-reducing layer hybridization among the conduction band states around $Q$ and $Q'$ and the valence band states in the center of the Brillouin zone at $\Gamma$. In contrast, hybridization is less effective in AA-stacked regions, upshifting the energies of $K\Gamma$ and $Q'\Gamma$ manifolds away from $KK$ interlayer and towards intralayer excitons of MoSe$_2$ and WSe$_2$ (two highest-energy states in the bottom panel of Fig.~\ref{fig2}b). Remarkably, this setting predicts the contributions of zero-momentum and finite-momentum interlayer excitons to HBL PL to stem from different stackings and thus from spatially distinct reservoirs. Consequently, the two-dimensional landscape of lowest-energy moir\'e excitons would thus be shaped by momentum-direct and indirect states residing in spatially separated domains of different stackings.    

The scenario is more simple for HTL PL with finite-momentum excitons being lowest in energy. For the peaks of Fig.~\ref{fig4}c and d, the absolute values of $g$-factors of about $12$ take momentum-direct $KK$ interlayer excitons as well as momentum-indirect reservoirs $K\Gamma$ and $Q\Gamma$ out of the picture. Among the former, dipole-active $KK$ states disqualify due to their $g$-factor of $\sim 6$, and $KK$ spin-like (spin-unlike) configurations formed by the electron in the lower (upper) MoSe$_2$ layer with theoretical $g$-factors between $11$ and $13$, as well as the respective $K'K$ counterparts with similar $g$-factors, are dismissed due to higher energies and thus negligibly small exciton populations. The latter momentum-indirect $K\Gamma$ and $Q\Gamma$ states exhibit only small $g$-factors because of the vanishing valley Zeeman term in the $\Gamma$ valley. By exclusion, the experimentally observed $g$-factors identify spin-like $QK$ and spin-unlike $Q'K$ interlayer excitons with theoretical $g$-factors of $\sim 10$ and $14$ as the only reasonable sources for the HTL PL peaks in the form of phonon sidebands. 

From the perspective of moir\'e exciton energy landscape governed by interlayer hybridization, lowest-energy HBL and HTL excitons should not differ in their spin-valley composition. Our analysis, however, suggests spin-like $K\Gamma$ or spin-unlike $Q'\Gamma$ states in HBL, and spin-like $QK$ or spin-unlike $Q'K$ states in HTL as lowest-energy manifolds. This controversy indicates that effects beyond hybridization have to be taken into account: the interplay of laterally modulated strain in moir\'e landscapes with opposite energy shifts for $K$ versus $Q$ and $\Gamma$ valleys \cite{Aslan2020}, and the combined piezoelectric and ferroelectric effects in the order of tens of meV \cite{Enaldiev2020} acting differently on interlayer excitons of distinct spin-valley configurations can reorder the hierarchy of energetically proximal interlayer exciton states. 

In conclusion, our experimental and theoretical study of excitons in twisted MoSe$_2$-WSe$_2$ HBL and HTL of R-type registry promote a complex picture of HBL PL. It is consistent with radiative recombination of zero-momentum $KK$ interlayer excitons and phonon-assisted emission from momentum-indirect reservoirs residing in spatially distinct regions of high-symmetry stackings. In contrast, the emission from the respective HTL system is entirely governed by phonon-assisted decay of momentum-dark $QK$ or $Q'K$ interlayer excitons. We base our conclusions on extensive optical spectroscopy experiments and calculations of the band structures, exciton states and $g$-factors for MoSe$_2$-WSe$_2$ HBL and HTL close to R-type registry. On these specific realizations of MoSe$_2$-WSe$_2$ heterostacks, our results highlight the primary role of moir\'e-modulated interlayer hybridization for the relaxation and formation of excitons in twisted van der Waals heterostructures with increasing layer number and structural complexity. Despite the extensive work presented here, a complete understanding of the rich phenomena observed in TMD heterostructures of different registries and alignment angles will require more efforts in experiment and theory to include piezoelectric and ferroelectric effects as well as strain at a detailed, quantitative level, and in the presence of reconstruction effects.

\clearpage

\textbf{Methods:}\\
The field-effect device based on a MoSe$_2$-WSe$_2$ heterostructure was fabricated by hot pick-up technique \cite{hot-pick-up}. First, a layer of hBN was picked up, followed by MoSe$_2$ with ML and BL regions, a ML of WSe$_2$ and a capping layer of hBN. The heterostack was subsequently placed in contact to a gold electrode that was deposited on a silver-coated glass substrate with a SiO$_2$ capping layer of $60$~nm. PL and DR experiments were performed in a home-built cryogenic microscope. The sample was mounted on piezo-stepping and scanning units (attocube systems, ANPxy101, ANPz101 and ANSxy100) for positioning with respect to a low-temperature objective (attocube systems, LT-APO/LWD/NIR/0.63 or LT-APO/NIR/0.81). The microscope was placed in a dewar with an inert helium atmosphere at a pressure of $20$~mbar and immersed in liquid helium at $4.2$~K or operated at $3.2$~K in a closed-cycle cryostat (attocube systems, attoDRY1000) equipped with a solenoid for magnetic fields of up to $\pm 9$~T. DR experiments were performed with a wavelength-tunable supercontinuum laser (NKT, SuperK Extreme or SuperK Varia), also used for PL excitation around $633$ or $715$~nm with repetition rates down to $2$~MHz. For continuous-wave measurements, the PL was excited with a laser diode at $635$~nm or a HeNe laser, spectrally dispersed by a monochromator (Roper Scientific, Acton SP~2750, SP~2558 or Acton SpectraPro 300i) and recorded with a nitrogen-cooled silicon CCD (Roper Scientific, PyLoN or Spec-10:100BR) or thermo-electrically cooled CCD (Andor iDus). Time-resolved PL was detected with an avalanche photodiode (Excelitas SPCM-AQRH) and correlated with a single photon counting system (PicoQuant, PicoHarp 300).

\textbf{Data availability}:\\
The data that support the findings of this study are available from the corresponding authors upon reasonable request.

\clearpage

%

\clearpage

\textbf{Competing interests:}\\
The authors declare no competing interests.

\textbf{Acknowledgements:}\\
This research was funded by the European Research Council (ERC) under the Grant Agreement No.~772195 as well as the Deutsche Forschungsgemeinschaft (DFG, German Research Foundation) within the Priority Programme SPP~2244 ''2DMP'' and the Germany's Excellence Strategy EXC-2111-390814868. Theoretical work was financially supported by the Foundation for the Advancement of Theoretical Physics and Mathematics ''BASIS''. A.\,S.\,B. has received funding from the European Union's Framework Programme for Research and Innovation Horizon 2020 (2014--2020) under the Marie Sk{\l}odowska-Curie Grant Agreement No.~754388, and from LMU Munich's Institutional Strategy LMUexcellent within the framework of the German Excellence Initiative (No.~ZUK22). S.\,Yu.\,K. acknowledges support from the Austrian Science Fund (FWF) within the Lise Meitner Project No. M 2198-N30, and A.\,H. from the Center for NanoScience (CeNS) and the LMUinnovativ project Functional Nanosystems (FuNS). K.\,W. and T.\,T. acknowledge support from the Elemental Strategy Initiative conducted by the MEXT, Japan, Grant Number JPMXP0112101001, JSPS KAKENHI Grant Numbers JP20H00354 and the CREST(JPMJCR15F3), JST.

\textbf{Contributions:}\\
M.~F., J.~S., J.~F., and V.~F. fabricated samples with high-quality hBN provided by K.~W. and T.~T. and performed experiments; M.~F., A.~S.~B. and A.~H. analyzed the data; A.~S.~B. developed theoretical concepts and performed calculations; I.~A.~V., S.~Yu.~K. performed numerical calculations; M.~F., A.~S.~B. and A.~H. prepared the figures and wrote the manuscript. All authors commented on the manuscript.



\clearpage

\clearpage

\end{document}


\title{SUPPLEMENTARY INFORMATION: \\
Moir\'e excitons in \texorpdfstring{MoSe$_2$-WSe$_2$}{MoSe2-WSe2} heterobilayers and heterotrilayers}

\author{Michael F\"org}
\affiliation{Fakult\"at f\"ur Physik, Munich Quantum Center, and
  Center for NanoScience (CeNS), Ludwig-Maximilians-Universit\"at
  M\"unchen, Geschwister-Scholl-Platz 1, 80539 M\"unchen, Germany}

\author{Anvar~S.~Baimuratov}
\affiliation{Fakult\"at f\"ur Physik, Munich Quantum Center, and
  Center for NanoScience (CeNS), Ludwig-Maximilians-Universit\"at
  M\"unchen, Geschwister-Scholl-Platz 1, 80539 M\"unchen, Germany}

\author{Stanislav~Yu.~Kruchinin}
\affiliation{Center for Computational Materials Sciences, Faculty of Physics, University of Vienna, Sensengasse 8/12, 1090 Vienna, Austria}
\affiliation{Nuance Communications Austria GmbH, Technologiestraße 8, 1120 Wien}

\author{Ilia~A.~Vovk}
\affiliation{Center of Information Optical Technology, ITMO University, Saint Petersburg 197101, Russia}

\author{Johannes Scherzer}
\affiliation{Fakult\"at f\"ur Physik, Munich Quantum Center, and
  Center for NanoScience (CeNS), Ludwig-Maximilians-Universit\"at
  M\"unchen, Geschwister-Scholl-Platz 1, 80539 M\"unchen, Germany}

\author{Jonathan F\"orste}
\affiliation{Fakult\"at f\"ur Physik, Munich Quantum Center, and
  Center for NanoScience (CeNS), Ludwig-Maximilians-Universit\"at
  M\"unchen, Geschwister-Scholl-Platz 1, 80539 M\"unchen, Germany}

\author{Victor Funk}
\affiliation{Fakult\"at f\"ur Physik, Munich Quantum Center, and
  Center for NanoScience (CeNS), Ludwig-Maximilians-Universit\"at
  M\"unchen, Geschwister-Scholl-Platz 1, 80539 M\"unchen, Germany}

\author{Kenji Watanabe}
\affiliation{Research Center for Functional Materials, National Institute for Materials Science, 1-1 Namiki, Tsukuba 305-0044, Japan}

\author{Takashi Taniguchi}
\affiliation{International Center for Materials Nanoarchitectonics, 
National Institute for Materials Science, 1-1 Namiki, Tsukuba 305-0044, Japan}

\author{Alexander H\"ogele}
\affiliation{Fakult\"at f\"ur Physik, Munich Quantum Center, and
  Center for NanoScience (CeNS), Ludwig-Maximilians-Universit\"at
  M\"unchen, Geschwister-Scholl-Platz 1, 80539 M\"unchen, Germany}
\affiliation{Munich Center for Quantum Science and Technology (MCQST),
  Schellingtra\ss{}e 4, 80799 M\"unchen, Germany}

\date{}

\maketitle


\noindent \textbf{Supplementary Note $\mathbf{1}$: Field effect characteristics of MoSe$_2$-WSe$_2$ heterostacks}

\vspace{11pt}

The field effect MoSe$_2$-WSe$_2$ heterostructure shown in Supplementary Figure~\ref{fig_sample} was fabricated by standard mechanical exfoliation using the hot pick-up technique \cite{hot-pick-up}. First, a layer of high quality hexagonal boron nitride (hBN) 
was picked up with the stamp, followed by MoSe$_2$ with monolayer (ML) and bilayer (BL) regions, a ML of WSe$_2$ and a capping layer of hBN. The whole stack was subsequently deposited on a target substrate consisting of a silver (Ag) electrode with a protective layer of $60$~nm SiO$_2$. The doping level in the entire MoSe$_2$-WSe$_2$ heterostructure was controlled by the gate voltage ($V_g$) applied to the Ag electrode in reference to a grounded gold electrode (Au) in contact with the WSe$_2$ ML. 

\begin{figure}[b!]
\centering
\includegraphics[scale=1.0]{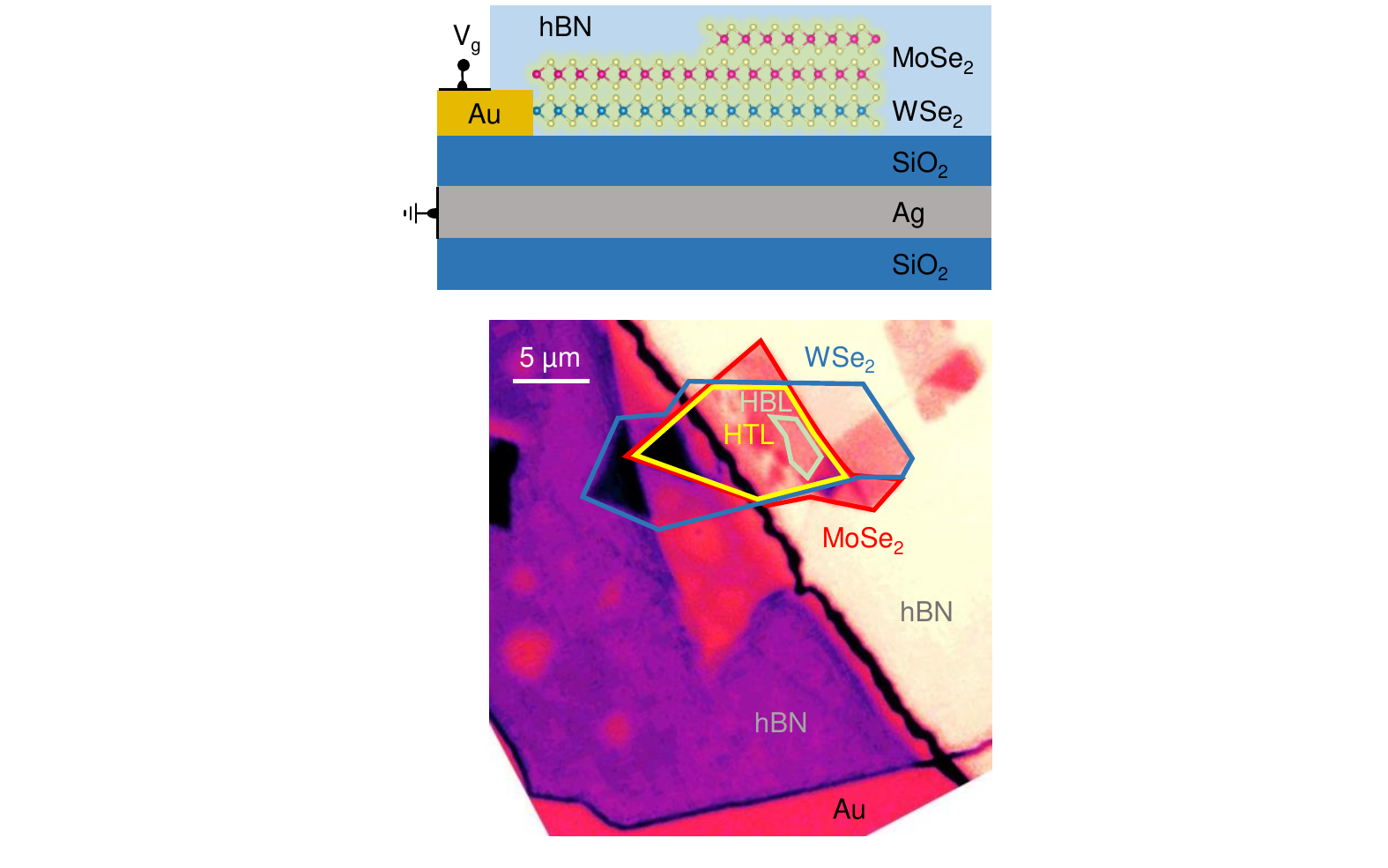}
\caption{Schematic layout (top) and optical micrograph (bottom) of the of the field effect heterostructure. MoSe$_2$ and WSe$_2$ few-layer crystals with monolayer and bilayer regions are delimited in red and blue, the resulting MoSe$_2$-WSe$_2$ HBL and HTL regions in light green and yellow, respectively.} \label{fig_sample}
\end{figure}

\begin{figure}[b!]
\centering
\includegraphics[scale=0.92]{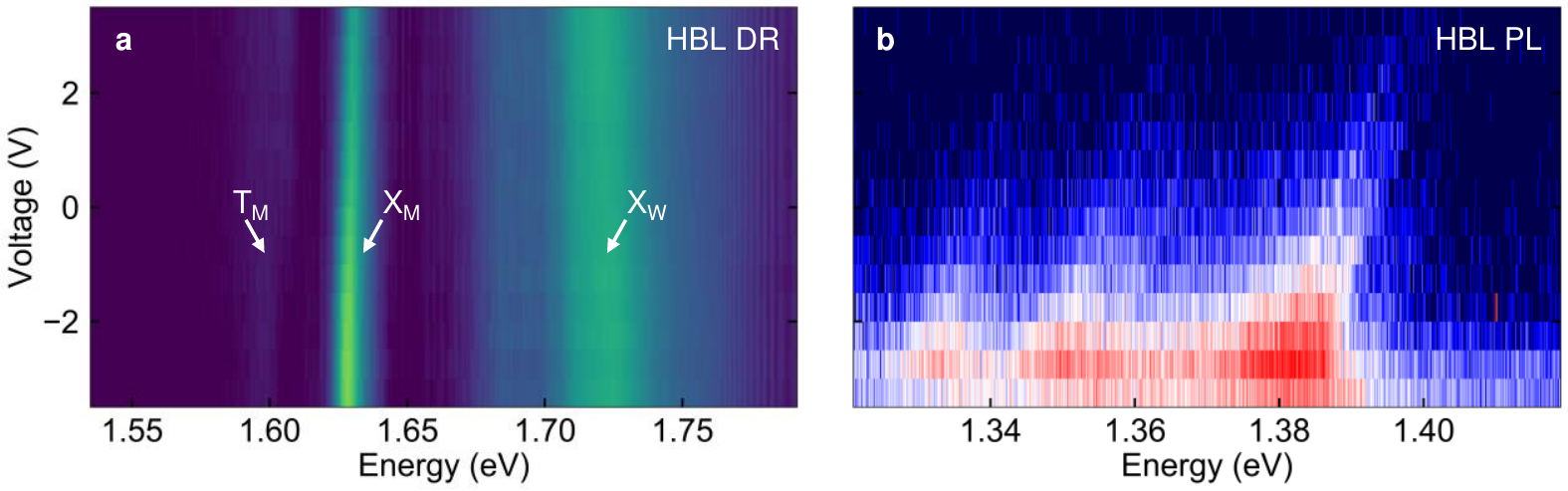}
\caption{\textbf{a}, Differential reflectivity (DR) of HBL in the spectral band of intralayer excitons as a function of gate voltage. The labelled features correspond to neutral excitons in ML MoSe$_2$ and WSe$_2$ (X$_M$ and X$_W$, respectively) and a vanishingly weak ML MoSe$_2$ trion (T$_M$). \textbf{b}, Interlayer exciton PL recorded on the same HBL position. Note that the gate voltage does not change the overall spectral shape.} \label{fig_dopingHBL}
\end{figure}

The charge carrier doping was determined from voltage-dependent differential reflectivity (DR) of intralayer exciton transitions in HBL (Supplementary Figure~\ref{fig_dopingHBL}a) and HTL (Supplementary Figure~\ref{fig_dopingHTL}a). The corresponding voltage-dependent PL from interlayer excitons is shown in Supplementary Figure~\ref{fig_dopingHBL}b and Supplementary Figure~\ref{fig_dopingHTL}b, respectively. 
The dominant MoSe$_2$ and WSe$_2$ ML features in the DR spectra of both Supplementary Figure~\ref{fig_dopingHTL}a and Supplementary Figure~\ref{fig_dopingHBL}a are related to the neutral intralayer exciton transitions X$_M$ and X$_W$, whereas the trion feature T$_M$ in ML MoSe$_2$ is only weakly expressed. This observation characterizes the heterostacks as very close to the charge neutrality condition throughout the accessible voltage interval explored experimentally. Note that although the peak emission energy and intensity depend on the gate voltage, the overall spectral structure of the interlayer PL exhibits no significant changes. All data of the main text were recorded at $0$~V. 


\begin{figure}[b!]
\centering
\includegraphics[scale=0.92]{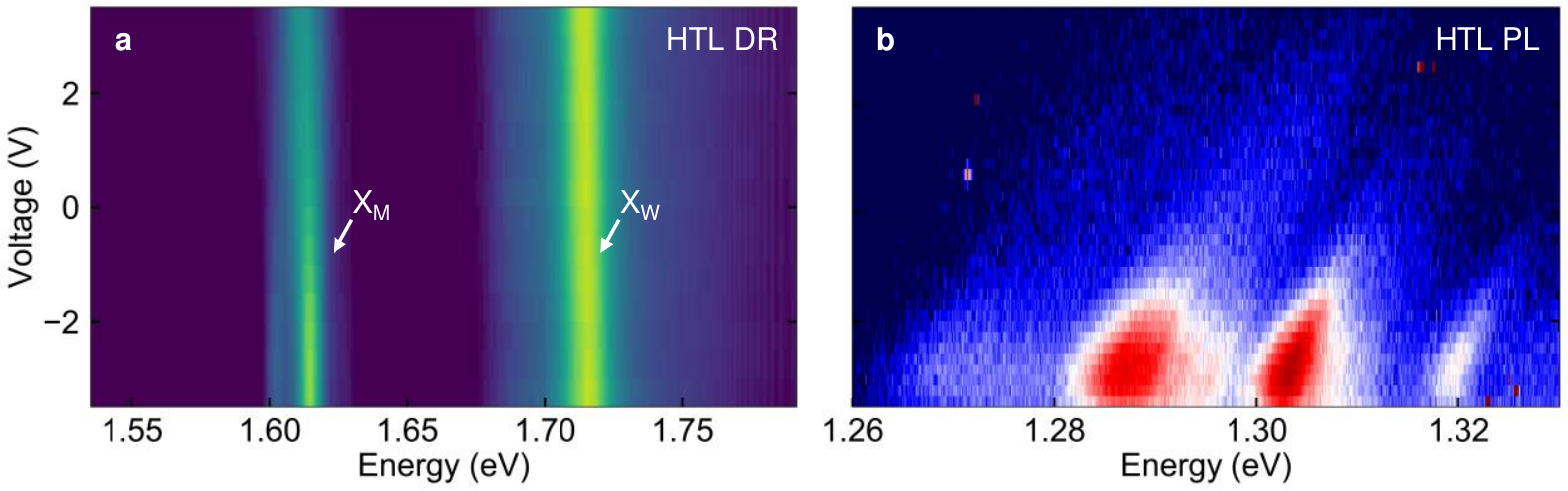}
\caption{\textbf{a}, Differential reflectivity (DR) of HTL in the spectral band of intralayer excitons as a function of gate voltage. The labelled features correspond to neutral excitons in ML MoSe$_2$ and WSe$_2$ (X$_M$ and X$_W$, respectively). \textbf{b}, Interlayer exciton PL recorded on the same HTL position. Note that the gate voltage does not change the overall spectral shape.} \label{fig_dopingHTL}
\end{figure}

%

\clearpage

\noindent \textbf{Supplementary Note $\mathbf{2}$: Photoluminescence spectra as a function of lateral displacement from MoSe$_2$-WSe$_2$ heterobilayer to heterotrilayer}

\begin{figure}[h!]
\includegraphics[scale=1.05]{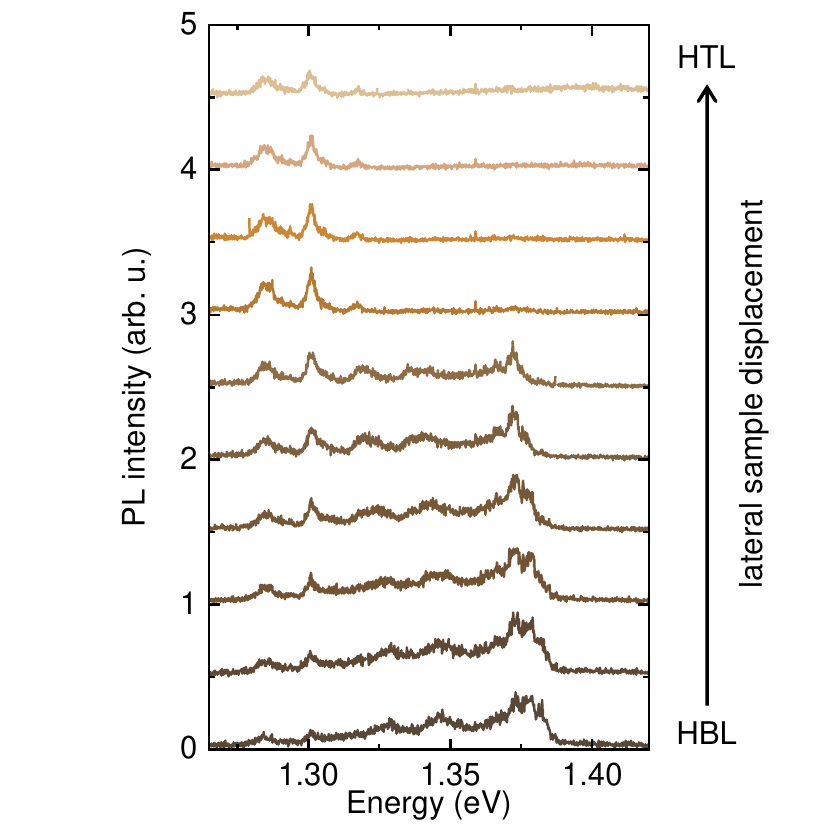}
\vspace{-12pt}
\caption{PL spectra upon lateral sample displacement from the HBL (bottom) to the HTL region (top).} \label{fig_transition}
\end{figure}

\vspace{11pt}
To confirm the reproducibility of the observations as a function of cooling cycles, the sample has been cool-cycled twice, using a low-temperature apochromatic objective with numerical aperture (NA) of $0.63$ (with the respective focal spot diameter of $\sim 1.5~\mu$m) in the first run, and $\mathrm{NA}=0.81$ (with a smaller spot diameter of $\sim 1.0~\mu$m) in the second run. All data shown in the main text were recorded with the higher-NA objective. Supplementary Figure~\ref{fig_transition}, with the data from the first run, exemplifies the dependence of the PL spectra on the lateral position. In this experiment, the sample was displaced with respect to fixed confocal excitation and detection spots by moving the heterostructure from the HBL to the HTL region. Evidently, the main features of HBL PL remain robust (peaks in the range $1.32 - 1.39$~eV in the bottom six spectra) before they entirely disappear on the HTL region where red-shifted peaks of HTL PL in the range $1.27 - 1.32$~eV (top four spectra) dominate. The weak yet finite PL cross-talk of HTL features into HBL spectra (that is further suppressed in Fig.~1 of the main text due to the smaller collection spot of the higher-NA objective) stems from insufficient lateral delimitation of the two regions with the objective of lower-NA.

\clearpage

\noindent \textbf{Supplementary Note $\mathbf{3}$: Dependence of photoluminescence multi-peak characteristics on gate voltage and excitation power}

\vspace{11pt}

As discussed in the context of Supplementary Figure~\ref{fig_dopingHTL} and Supplementary Figure~\ref{fig_dopingHBL} the multi-peak PL structure is overall robust within the experimentally explored gate voltage range. The quantitative analysis of the spectra as a function of gate voltage recorded on a heterostructure region with contributions from both HBL and HTL stacks (as in the bottom spectrum of Supplementary Figure~\ref{fig_transition}) is shown in Supplementary Figure~\ref{fig_peaks}. From the data in Supplementary Figure~\ref{fig_peaks}a it is evident that all peaks exhibit similar intensity variations, whereas the data in Supplementary Figure~\ref{fig_peaks}b and c confirm that the multi-peak PL structure is preserved throughout the gate voltage range between $-1.5$ and $1.5$~V. In particular, the energy difference of $30$~meV between the blue-most peaks (P1 and P2) and $15$~meV between all other consecutive peaks is constant in the entire gate voltage range.    

\begin{figure}[h!]
\includegraphics[scale=0.96]{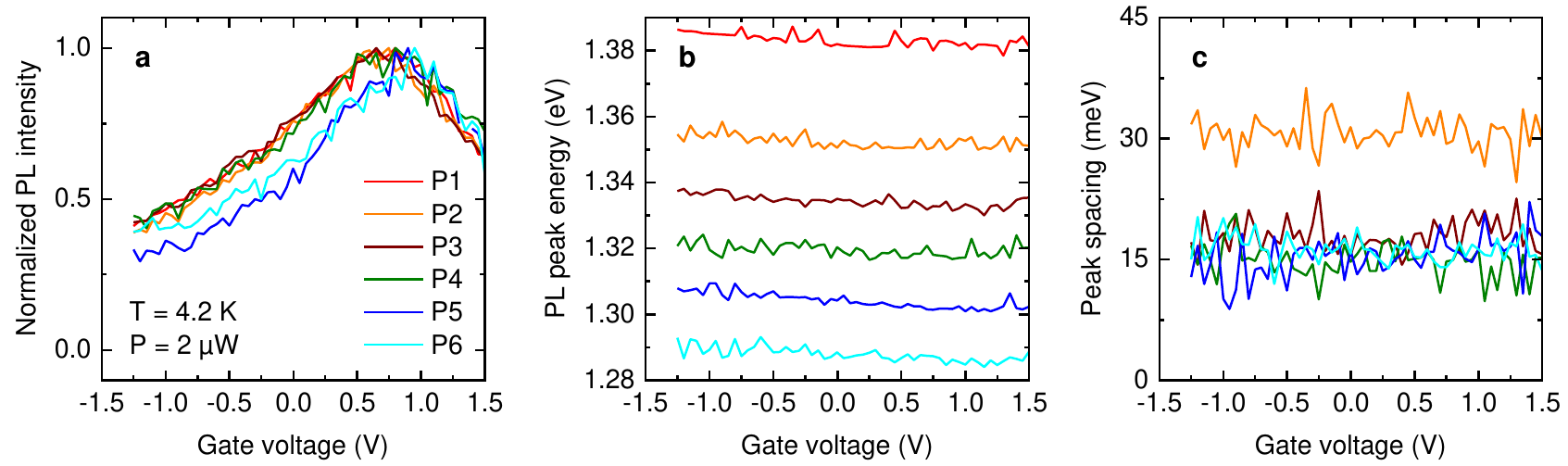}
\caption{\textbf{a} and \textbf{b},  Photoluminescence intensity and energy of peaks in the HBL and HTL spectra recorded on a sample position with contributions from both heterostacks. The peaks are numbered from P1 to P6 with decreasing emission energy. \textbf{c}, Energy difference between consecutive peaks as a function of gate voltage.} \label{fig_peaks}
\end{figure}

Qualitatively similar trends were observed as a function of excitation power in Supplementary Figure~\ref{fig_HBL-HTL-power}. The left and right panels of Supplementary Figure~\ref{fig_HBL-HTL-power} show voltage-dependent HBL (the red-most peaks are due to cross-talk from the HTL region into the collection spot) and HTL PL for various laser excitation powers. Apart from power-dependent screening the set of data demonstrates that the characteristic multi-peak PL structures of both HBL and HTL are preserved over two orders of magnitude in excitation power. In particular, the intensity of the highest-energy HBL peak is highest at all powers, and the ratio to other peaks does not reverse down to the lowest excitation power of $0.3~\mu$W.

\begin{figure}[t!]
\vspace{-10pt}
\includegraphics[scale=1.0]{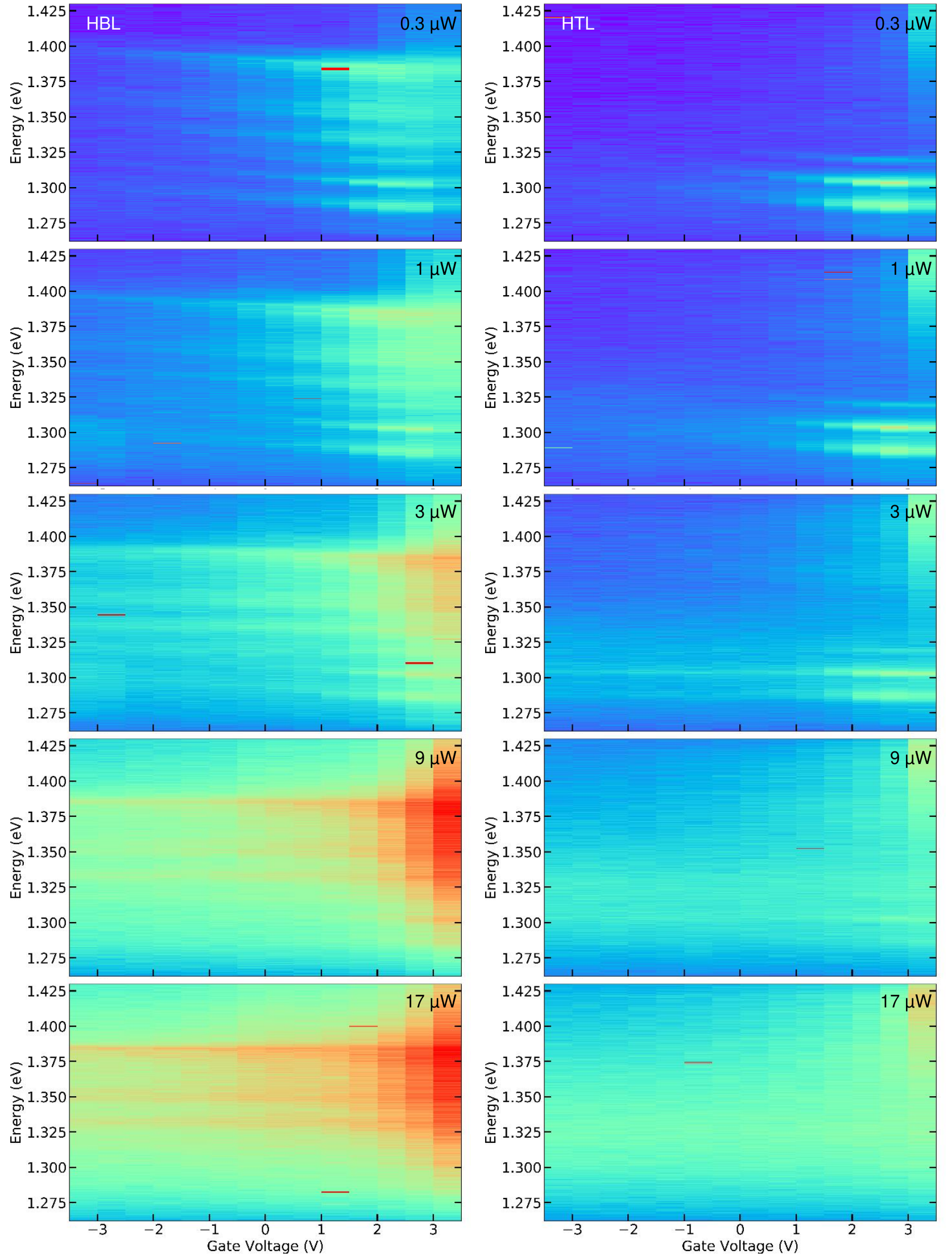}
\vspace{-40pt}
\caption{HBL (left) and HTL (right) photoluminescence as a function of gate voltage at different excitation powers given explicitly in each graph.} \label{fig_HBL-HTL-power}
\end{figure}

\clearpage

\noindent \textbf{Supplementary Note $\mathbf{4}$: Photoluminescence spectra of MoSe$_2$-WSe$_2$ heterotrilayer and native WSe$_2$ homobilayer}

\vspace{11pt}
A striking similarity in the PL spectra from MoSe$_2$-WSe$_2$ HTL and native WSe$_2$ BL is evident from Supplementary Figure~\ref{fig_compare}. Native WSe$_2$ BL exhibits PL emission as phonon-sidebands of momentum indirect $QK$ excitons \cite{Lindlau2017BL}. The spectra of MoSe$_2$-WSe$_2$ HTL and WSe$_2$ BL feature remarkably similar profiles when red-shifted by the band-offset energy of $\sim 280$~meV.

\begin{figure}[h!]
\includegraphics[scale=1.0]{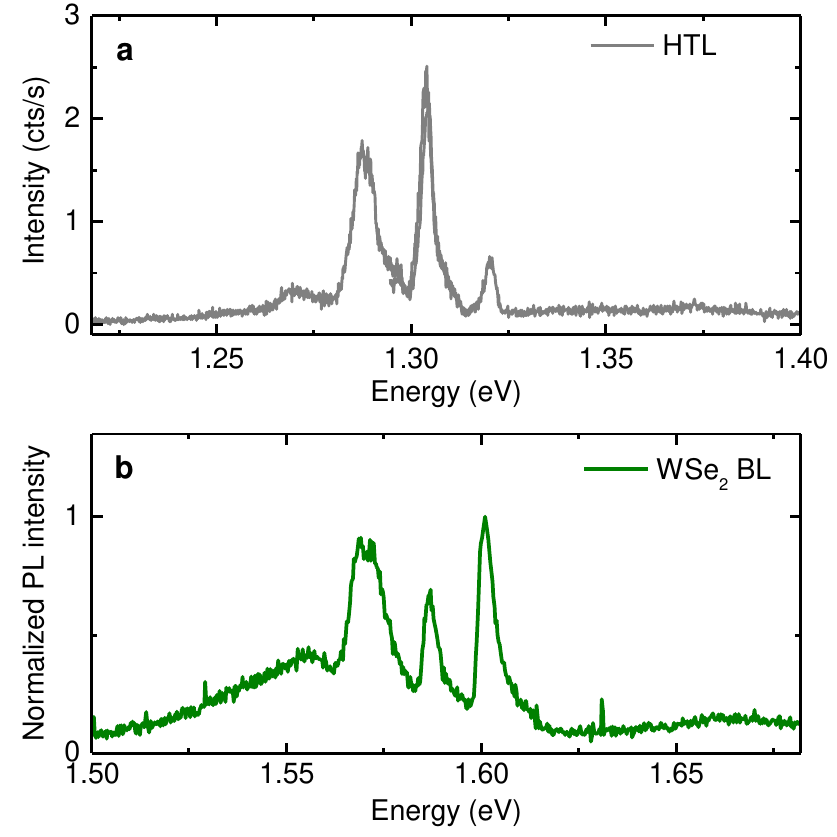}
\vspace{-12pt}
\caption{\textbf{a}, Photoluminescence spectrum of interlayer excitons in MoSe$_2$-WSe$_2$ HTL and \textbf{b}, native WSe$_2$ BL encapsulated in hBN. For both measurements the temperature was $3.2$~K.} \label{fig_compare}
\end{figure}


\noindent \textbf{Supplementary Note $\mathbf{5}$: Time-resolved photoluminescence decay}

\vspace{11pt}

\begin{figure}[!ht]
\includegraphics[scale=1.04]{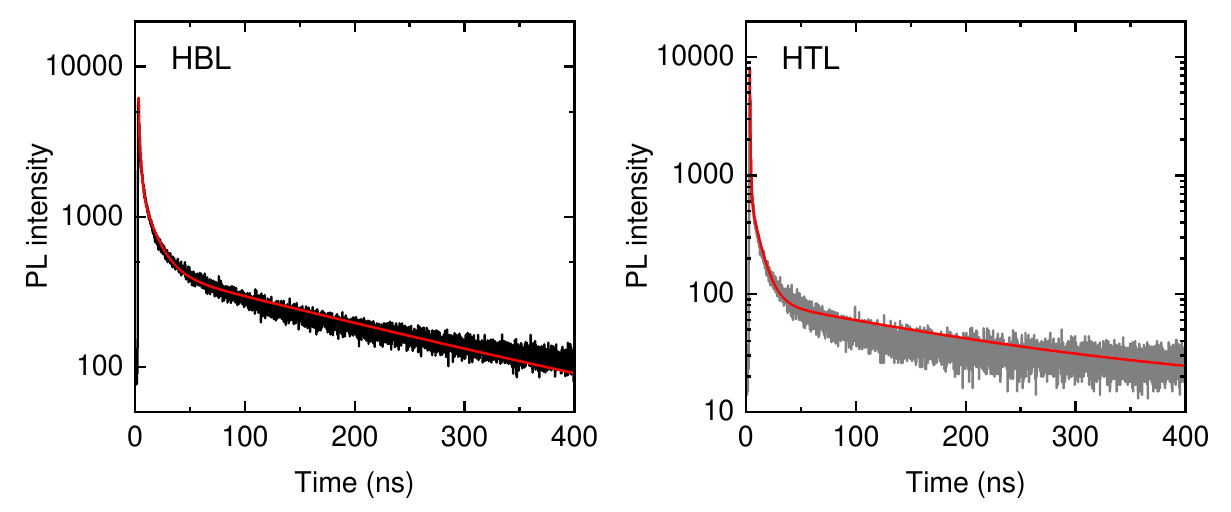}
\vspace{-12pt}
\caption{Photoluminescence decay in MoSe$_2$-WSe$_2$ HBL (left) and HTL (right) at zero gate voltage. The excitation was performed at $633$~nm and $1~\mu$W power in the focal spot. The solid red lines show tri-exponential decay as best fits.} \label{fig_decay}
\end{figure}

The PL dynamics of HBL and HTL PL were studied with time-resolved PL spectroscopy within various spectral bands. As evident from Supplementary Figure~\ref{fig_decay} and Supplementary Figure~\ref{fig_compare}, the spectrally structured interlayer exciton emission of HBL and HTL exhibited similar PL decay characteristics. Good approximation to the PL decay was obtained with three exponential decay channels with lifetimes of $3$, $12$ and $480$~ns for HBL and $1$, $12$ and $300$~ns for HTL emission. Remarkably, the contributions of the individual decay channels to the total PL decay varied only marginally across the spectral bands. For both sample regions, the decay was dominated by the slow decay component (with a weight of $89$ and $80\%$ in HBL and HTL, respectively) with contributions of the intermediate and fast decay channels of $\sim 10\%$ (with weights of $8$ and $13\%$ for the intermediate and $3$ and $7\%$ for the rapid components of HBL and HTL PL decay, respectively).

In Supplementary Figure~\ref{fig_compare}, we summarize the decay data recorded in different spectral bands as indicated by colored regions in the top and bottom panels of Supplementary Figure~\ref{fig_compare}a for HBL and HTL peaks, respectively. The decay times and weights of the three characteristic decay channels, obtained from tri-exponential decay fits within each spectral band, are summarized in Supplementary Figure~\ref{fig_compare}b and c, respectively. The data suggest different PL decay characteristics within HBL and HTL regions with only little dependence on the emission energy for a given heterostack. This observation is inconsistent with lower PL peaks originating from exciton-localizing defects. 

\begin{figure}[!ht]
\includegraphics[scale=1.04]{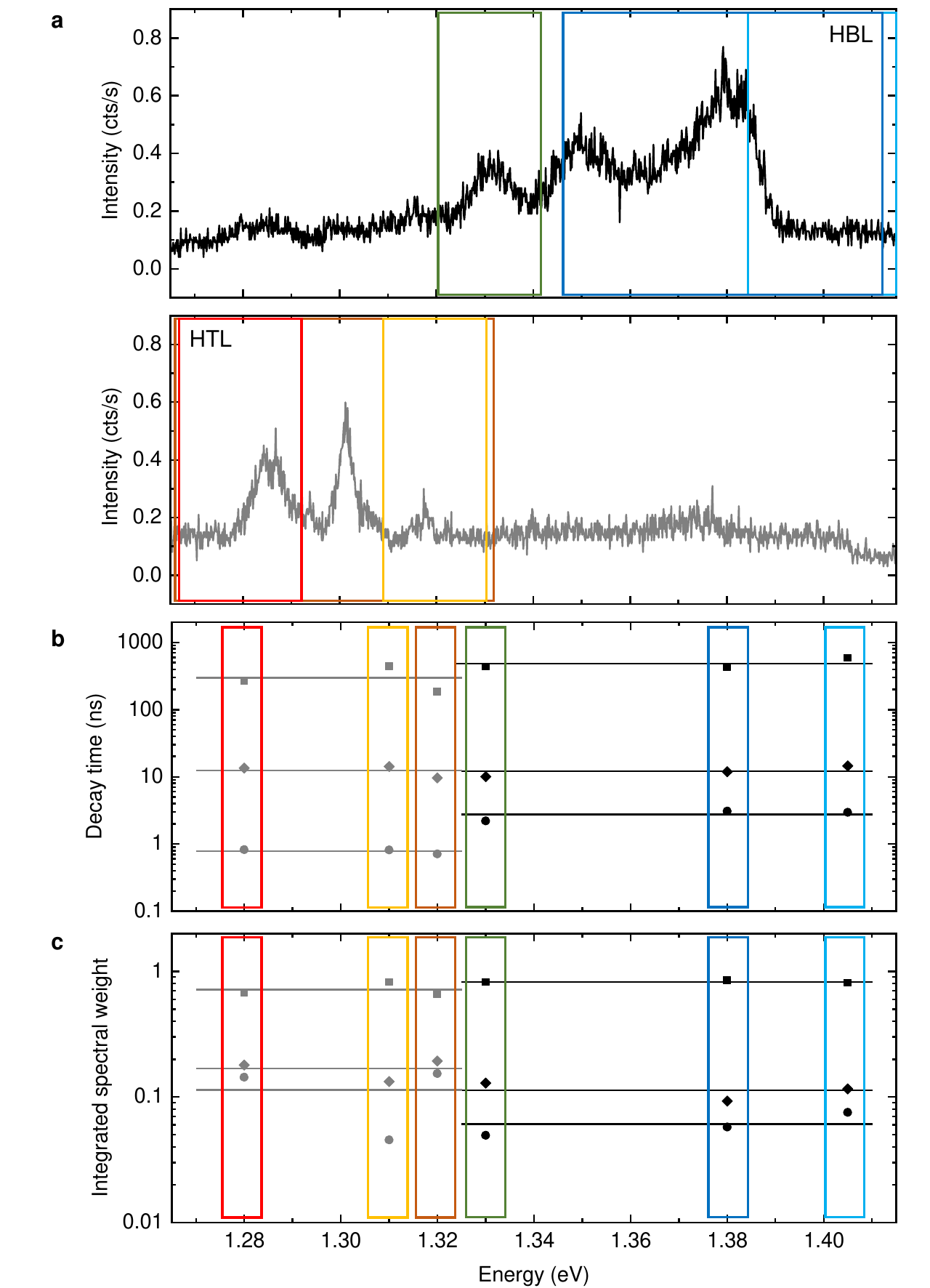}
\vspace{-12pt}
\caption{\textbf{a}, Photoluminescence spectra of MoSe$_2$-WSe$_2$ HBL (top) and HTL (bottom) with colored spectral windows used to record time-resolved PL decay. \textbf{b} and \textbf{c}, Decay times and integrated spectral weights for three decay channels of the tri-exponential PL decay obtained from fits to spectrally limited PL bands of HBL (black data points, average values showed by black solid lines) and HTL (grey data points, average values showed by grey solid lines).} \label{fig_compare}
\end{figure}

\clearpage

\noindent \textbf{Supplementary Note $\mathbf{6}$: Exciton energies in MoSe$_2$-WSe$_2$ heterobilayers and heterotrilayers}
\vspace{11pt}

Density functional theory (DFT) calculations of MoSe$_2$-WSe$_2$ HBL and HTL were performed with the PBEsol exchange-correlation functional \cite{Csonka_2009_PRB_79_155107} as implemented in the Vienna ab initio simulation package (VASP) \cite{Shishkin_2007_PRB_75_235102}. Van der Waals interactions were included with the DFT-D3 method by \textcite{Grimme_2010_JCP_132_154104} with Becke-Johnson damping \cite{Grimme_2011_JCC_32_1456}. Moreover, spin-orbit interactions were included at all stages. Elementary cells with thickness of 35~\AA{} in the $z$-direction were used in order to minimize interactions between periodic images. The atomic positions were relaxed with a cutoff energy of $400$~eV until the total energy change was less than $10^{-6}$~eV. Calculations were performed for high-symmetry points of HBL and HTL moir\'e patterns in R-type stacking as shown in Supplementary Figure~\ref{Stacking} on the $\Gamma$-centered $\mathbf k$ grid of $6 \times 6$ divisions with the cutoff energy of $300$~eV, with 600 bands for the HBL and 900 bands for the HTL structures. The results for energy gaps and effective masses (in units of free electron mass $m_0$) are summarized in Supplementary Tables~\ref{tab01} and \ref{tab02} for HBL and HTL.

\begin{figure}[h!]
  \includegraphics[scale=0.9]{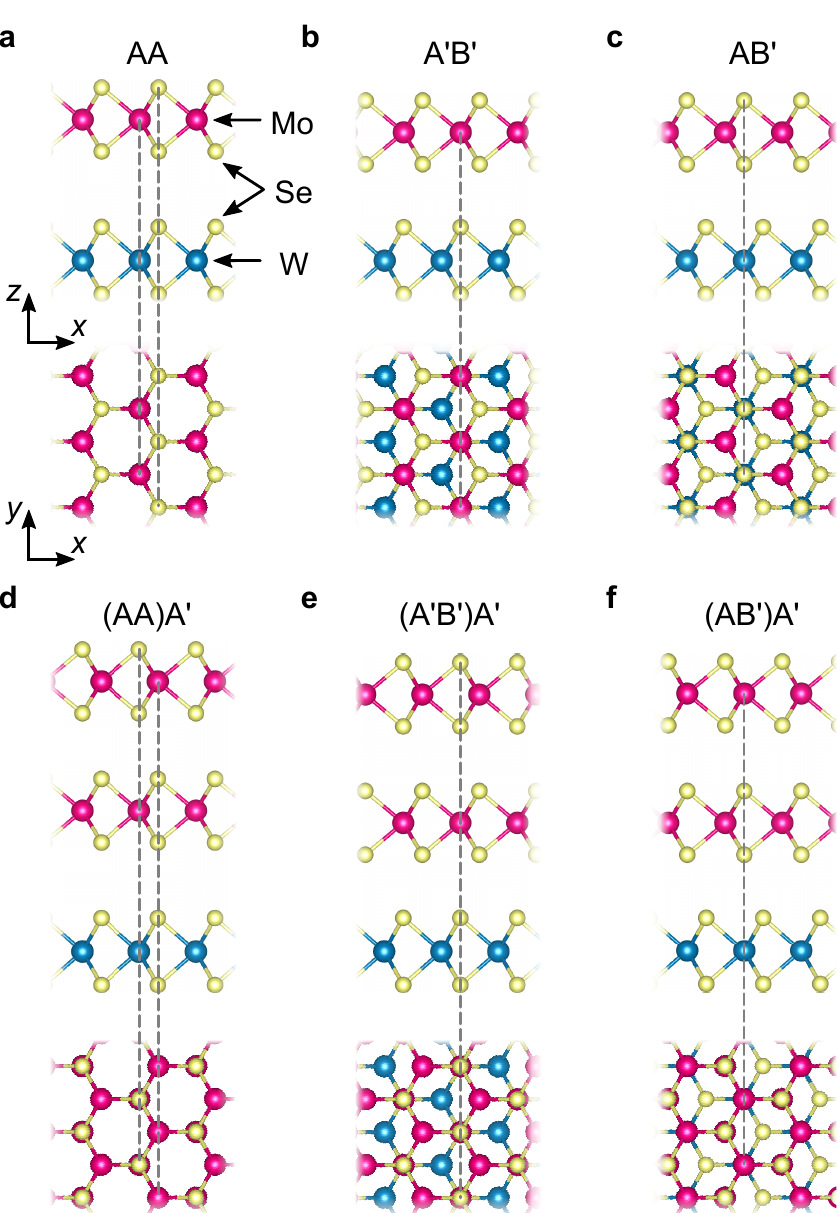}
  \vspace{-5pt}
  \caption{High-symmetry stackings in R-type HBL (\textbf{a--c}) and HTL (\textbf{d--f}) MoSe$_2$-WSe$_2$.}
  \label{Stacking}
\end{figure}

\begin{table}[tb]
  \caption{Effective masses and energy gaps of HBL MoSe$_2$-WSe$_2$ from DFT.}
  \begin{ruledtabular}
\begin{tabular}{c|ccC|ccC|C}
& \multicolumn{3}{c|}{Electron} & \multicolumn{3}{c|}{Hole} & \textrm{Energy gap} \\
Stacking & k-point & Layer & \me (m_0)  & k-point & Layer & \mh (m_0) & E_\text{g}~\text{(eV)} \\
    \hline
    AA       & K       & W     & 0.38       & K       & W     & 0.47      & 1.4   \\
             & K       & Mo    & 0.67       & K       & Mo    & 0.74      & 1.43  \\
             & K       & Mo    & 0.7        & K       & W     & 0.47      & 1.12  \\
             & K       & Mo    & 0.67       & K       & W     & 0.47      & 1.1   \\
             & Q       & Mo/W  & 0.66       & K       & W     & 0.47      & 1.05  \\
             & K       & Mo    & 0.67       & $\Gamma$& W/Mo  & 1.36      & 1.43  \\
             & Q       & Mo/W  & 0.66       & $\Gamma$& W/Mo  & 1.36      & 1.38  \\
    \hline
    A'B'     & K       & W     & 0.37       & K       & W     & 0.44      & 1.39  \\
             & K       & Mo    & 0.63       & K       & Mo    & 0.71      & 1.42  \\
             & K       & Mo    & 0.71       & K       & W     & 0.44      & 1.06  \\
             & K       & Mo    & 0.63       & K       & W     & 0.44      & 1.04  \\
             & Q       & Mo/W  & 0.66       & K       & W     & 0.44      & 1.01  \\
             & K       & Mo    & 0.63       & $\Gamma$& W/Mo  & 0.85      & 1.09  \\
             & Q       & Mo/W  & 0.66       & $\Gamma$& W/Mo  & 0.85      & 1.06  \\
    \hline
    AB'      & K       & W     & 0.39       & K       & W     & 0.46      & 1.4   \\
             & K       & Mo    & 0.66       & K       & Mo    & 0.72      & 1.41  \\
             & K       & Mo    & 0.73       & K       & W     & 0.46      & 1.2   \\
             & K       & Mo    & 0.66       & K       & W     & 0.46      & 1.18  \\
             & Q       & Mo/W  & 0.61       & K       & W     & 0.46      & 1.08  \\  
             & K       & Mo    & 0.66       & $\Gamma$& W/Mo  & 0.92      & 1.20  \\
             & Q       & Mo/W  & 0.61       & $\Gamma$& W/Mo  & 0.92      & 1.10
  \end{tabular}
  \end{ruledtabular}
  \label{tab01}
\end{table}

\begin{table}[tb]
\caption{Effective masses and energy gaps of HTL MoSe$_2$-WSe$_2$ from DFT.}
\begin{ruledtabular}
\begin{tabular}{c|ccC|ccC|C}

& \multicolumn{3}{c|}{Electron} & \multicolumn{3}{c|}{Hole} & \textrm{Energy gap} \\
Stacking & k-point & Layer    & \me (m_0)  & k-point & Layer & \mh (m_0) & E_\text{g}~\text{(eV)} \\
    \hline
    (AA)A'   & K       & W        & 0.39       & K       & W     & 0.48      & 1.4                              \\
             & K       & Mo'      & 0.83       & K       & Mo'   & 0.79      & 1.42                             \\
             & K       & Mo       & 0.82       & K       & Mo    & 0.74      & 1.42                             \\
             & K       & Mo'      & 0.58       & K       & W     & 0.48      & 1.11                             \\
             & K       & Mo       & 0.6        & K       & W     & 0.48      & 1.1                              \\
             & K       & Mo'      & 0.83       & K       & W     & 0.48      & 1.09                             \\
             & K       & Mo       & 0.82       & K       & W     & 0.48      & 1.08                             \\
             & Q       & Mo/Mo'/W & 0.57       & K       & W     & 0.48      & 0.88                             \\
             & K       & Mo       & 0.82       & $\Gamma$& Mo/Mo'/W  & 0.80      & 1.21                         \\
             & Q       & Mo/Mo'/W & 0.57       & $\Gamma$& Mo/Mo'/W  & 0.80      & 1.01                         \\
    \hline
    (A'B')A' & K       & W        & 0.39       & K       & W     & 0.46      & 1.39                             \\
             & K       & Mo'      & 0.83       & K       & Mo'   & 0.75      & 1.42                             \\
             & K       & Mo       & 0.81       & K       & Mo    & 0.74      & 1.42                             \\
             & K       & Mo'      & 0.61       & K       & W     & 0.46      & 1.07                             \\
             & K       & Mo       & 0.58       & K       & W     & 0.46      & 1.06                             \\
             & K       & Mo'      & 0.83       & K       & W     & 0.46      & 1.05                             \\
             & K       & Mo       & 0.81       & K       & W     & 0.46      & 1.04                             \\
             & Q       & Mo/Mo'/W & 0.53       & K       & W     & 0.46      & 0.83                             \\  
             & K       & Mo       & 0.81       & $\Gamma$& Mo/W/Mo'  & 0.94      & 1.09                         \\
             & Q       & Mo/Mo'/W & 0.53       & $\Gamma$& Mo/W/Mo'  & 0.94      & 0.88                         \\
    \hline
    (AB')A'  & K       & W        & 0.38       & K       & W     & 0.44      & 1.4                              \\
             & K       & Mo'      & 0.81       & K       & Mo'   & 0.69      & 1.42                             \\
             & K       & Mo       & 0.79       & K       & Mo    & 0.73      & 1.4                              \\
             & K       & Mo'      & 0.63       & K       & W     & 0.44      & 1.21                             \\
             & K       & Mo       & 0.56       & K       & W     & 0.44      & 1.2                              \\
             & K       & Mo'      & 0.81       & K       & W     & 0.44      & 1.19                             \\
             & K       & Mo       & 0.79       & K       & W     & 0.44      & 1.18                             \\
             & Q       & Mo/Mo'/W & 0.55       & K       & W     & 0.44      & 0.95                             \\
             & K       & Mo       & 0.79       & $\Gamma$& Mo/W/Mo'  & 0.82      & 1.11                         \\
             & Q       & Mo/Mo'/W & 0.55       & $\Gamma$& Mo/W/Mo'  & 0.82      & 0.89                                       
  \end{tabular}
  \end{ruledtabular}
  \label{tab02}
\end{table}

Based on these DFT results, we used the Wannier exciton model in the effective mass approximation \cite{Berghaeuser2014} to calculate the exciton energies $E_X$ for different spin-valley configurations shown in Fig.~2 of the main text, obtained as $E_X = E_\text{g}^\text{DFT} + E_\text{g}^\text{offset} - E_\text{b}$ from the DFT quasiparticle band gap energy $E_\text{g}^\text{DFT}$ corrected by an offset $E_\text{g}^\text{offset}$ that accounts for an underestimated band gap, and the exciton binding energy $E_\text{b}$. $E_\text{g}^\text{offset} = 480$~meV was used as a global energy offset for all exciton configurations by placing experimental and theoretical energy positions of $KK$ intralayer excitons of WSe$_2$ in resonance. 

To determine the exciton binding energy we solved the stationary Schr\"odinger equation of the electron-hole relative motion,
\begin{equation}
  \left[ -\frac{\hbar^2}{2\mu} \nabla^2 + V(\rho) \right] \psi(\rho) =
    E_n \psi(\rho),
\end{equation}
where $\psi (\rho)$ is the radial wave function, $\mu = \me \mh /(\me + \mh)$ is the reduced effective mass, $\me$ and $\mh$ are the effective masses of electron and hole, and $V(\rho)$ the Rytova--Keldysh potential \cite{Keldysh1979,Rytova1967} of the form
\vspace{-9pt}
\begin{equation}
    V(\rho) =
    -\frac{\pi e^2}{2 \varepsilon \rho_0}
    \left[
        H_0 \left( \frac{\rho}{\rho_0} \right) 
        - Y_0 \left( \frac{\rho}{\rho_0} \right) 
    \right],
\end{equation}
with elementary charge $e$, effective dielectric constant $\varepsilon$, screening length $\rho_0$, and Struve and Neumann functions $H_0 (x)$ and $Y_0 (x)$. The binding energy was obtained as $E_\text{b} = -\text{min} (E_n)$ from the minimal eigenvalue $E_n$, with $\varepsilon = 4.5$ and $\rho_0 = 1$~nm as parameters for MoSe$_2$ \citep{Han2018}. Remarkably, this procedure of calculating the exciton energy based on the Wannier model and DFT data yields good quantitative agreement with the computationally expensive simulations based on many-body approximations in HBL case \cite{Gillen2018}. However, different effective screening in the Coulomb interactions among electrons and holes in different valleys would result in relative energy shifts. A difference of $\varepsilon$ by $10\%$ for $KK$ and $QK$ states, for example, would yield an energy shift of about $35$~meV. Presently, this uncertainty in the relative energy positions of interlayer exciton states in different valley configurations can not be eliminated from our theory as the valley-dependence of the effective dielectric screening is quantitatively unknown.

\vspace{11pt}
\noindent \textbf{Supplementary Note $\mathbf{7}$: Calculation of exciton $g$-factors}
\vspace{4pt}

Our methodology for calculations of exciton $g$-factors from first principles has been described in detail in \cite{Foerste2020}. Here, we recapitulate the main steps for determining the $g$-factors of excitons in different spin and valley configurations in HBL and HTL MoSe$_2$-WSe$_2$. The exciton is formed by Coulomb correlations between an occupied state in the conduction band $c$ with the wave vector $\mathbf k_c$ and spin $z$-projection $s_c$ and an empty state in the valence band $v$ with the wave vector $\mathbf k_v$ and spin $z$-projection $s_v$. In this spin-valley configuration, the exciton $g$-factor is given by  
\vspace{-11pt}
\begin{equation}
  g^{(cv)} (\mathbf k_c, \mathbf k_v) =
    g_c (\mathbf k_c) - g_v (\mathbf k_v),
\vspace{-11pt}
\end{equation}
\vspace{-11pt}
where the $g$-factor of the electron in band $n = c,v$ is 
\begin{equation}
  g_n (\mathbf k) =
    g_0 s_n + 2 L_n (\mathbf k).
\vspace{-11pt}
\end{equation}
Here, $g_0 = 2$ is the free electron Land\'e factor, and the $z$-component of the orbital angular momentum \cite{Roth1959,Bir1974,Xiao2010,Wang_2015} is
\vspace{-11pt}
\begin{equation} \label{eq_lz}
  L_n (\mathbf k) = \\
    \frac{2 m_0}{\hbar^2}
    \sum_{m \neq n}
      \textrm{Im}\left[\xi_{nm}^{(x)} (\mathbf k) \xi_{mn}^{(y)}(\mathbf k)\right]
      (E_{n \mathbf k} - E_{m \mathbf k}).
\vspace{-11pt}
\end{equation}
In the summation, the index $m$ runs over all bands excluding the band of interest, $\bm{\xi}_{nm} (\mathbf k) = i \bra{u_{n\mathbf k}} \partial / \partial \mathbf k \ket{u_{m\mathbf k}}$ is the interband matrix element of the coordinate operator, $E_{n \mathbf k}$ and $u_{n\mathbf k}$ are the energy and periodic Bloch amplitude of the electron in band $n$ with wave vector $\mathbf k$. 

Using the energy band structure and interband matrix elements of the coordinate operator obtained from DFT we calculate the $g$-factor for excitons in different spin and valley configurations in HBL and HTL MoSe$_2$-WSe$_2$ according to the equations above. The results are summarized in Supplementary Table~\ref{tab03} for momentum-direct $KK$ and momentum-indirect $KK'$, $QK$, $Q'K$, $K\Gamma$, and $K'\Gamma$ interlayer excitons in HBL (top block) and HTL (bottom blocks) in different stackings of R-type registry. 

\begin{table}[t!]
\caption{Calculated $g$-factors of interlayer excitons in R-type MoSe$_2$-WSe$_2$ HBL (topmost block) and HTL (bottom blocks) in spin-like ($\uparrow \uparrow$) and spin-unlike ($\downarrow \uparrow$) configurations of conduction band electrons in $K, K', Q$ or $Q'$ valleys of MoSe$_2$ and empty valence band states at $K$ in WSe$_2$ or at $\Gamma$ in the hybrid band of MoSe$_2$-WSe$_2$. For each spin-valley configuration, the $g$-factors corresponding to the lower-energy state are shown in bold. For HTL, the upper (lower) block shows $KK$, $K'K$, $K\Gamma$, and $K'\Gamma$ excitons with the conduction band electron localized in the lower (upper) MoSe$_2$ layer, as well as $QK$, $Q'K$, $Q\Gamma$, and $Q'\Gamma$ excitons with small (large) hybridization with WSe$_2$ conduction band states. The sign convention for $KK$ interlayer exciton $g$-factors is the same as for the $KK$ intralayer exciton in WSe$_2$; only absolute values are given for momentum-indirect interlayer excitons as well as for direct $KK$ excitons with $z$-polarized in-plane emission.}
  \begin{ruledtabular}
\begin{tabular}{c|RR|RR|RR|RR|RR|RR|RR|RR}
& \multicolumn{2}{c|}{$KK$} & \multicolumn{2}{c|}{$K'K$} & \multicolumn{2}{c|}{$QK$} & \multicolumn{2}{c|}{$Q'K$}& \multicolumn{2}{c|}{$K\Gamma$} & \multicolumn{2}{c|}{$K'\Gamma$} & \multicolumn{2}{c|}{$Q\Gamma$} & \multicolumn{2}{c}{$Q'\Gamma$}
\\
 Stacking & \uparrow \uparrow & \downarrow \uparrow & \uparrow \uparrow & \downarrow \uparrow & \uparrow \uparrow & \downarrow \uparrow & \uparrow \uparrow & \downarrow \uparrow & \uparrow \uparrow & \downarrow \uparrow & \uparrow \uparrow & \downarrow \uparrow & \uparrow \uparrow & \downarrow \uparrow & \uparrow \uparrow & \downarrow \uparrow
        \\
        \hline
       AA & \mathbf{-6.4} &  11.0          & 13.0         & \mathbf{17.6} & \mathbf{9.0} & 13.3 & 10.7 & \mathbf{15.0} & \mathbf{3.6} & 1.0          & 3.0          & \mathbf{7.6} & \mathbf{1.0} & 3.3 & 0.7 & \mathbf{5.0}  \\
     A'B' & \mathbf{+5.8} & -10.5          & 13.1         & \mathbf{17.8} & \mathbf{8.6} & 13.0 & 10.6 & \mathbf{14.9} & \mathbf{4.0} & 0.7          & 3.3          & \mathbf{8.0} & \mathbf{1.1} & 3.2 & 0.8 & \mathbf{5.1}  \\
      AB' &  \mathbf{6.3} & +10.9          & 13.0         & \mathbf{17.6} & \mathbf{8.7} & 12.9 & 11.0 & \mathbf{15.3} & \mathbf{3.7} & 1.0          & 3.0          & \mathbf{7.7} & \mathbf{1.3} & 2.9 & 1.1 & \mathbf{5.3}  \\
        \hline
   (AA)A' & \mathbf{-6.3} &  11.6          & 12.6         & \mathbf{17.9} & \mathbf{9.9} & 14.1 & 10.1 & \mathbf{14.3} & \mathbf{3.8} & 1.5          & 2.5          & \mathbf{7.8} & \textbf{0.2} & 4.0 & 0.0 & \textbf{4.2}  \\
 (A'B')A' & \mathbf{+5.9} & -10.8          & 13.1         & \mathbf{17.9} & \mathbf{9.5} & 13.8 & 10.1 & \mathbf{14.4} & \mathbf{4.0} & 0.8          & 3.2          & \mathbf{8.0} & \textbf{0.4} & 3.8 & 0.2 & \textbf{4.4}  \\
  (AB')A' &  \mathbf{6.3} & +12.2          & 12.0         & \mathbf{18.0} & \mathbf{9.6} & 13.9 & 10.4 & \mathbf{14.6} & \mathbf{3.8} & 2.1          & 1.9          & \mathbf{7.8} & \textbf{0.5} & 3.7 & 0.3 & \textbf{4.5}  \\
        \hline
   (AA)A' & +12.8         & \mathbf{-18.1} & \mathbf{6.1} & 11.4          & 9.1          & 13.3 & 10.9 & 15.1          & 2.7          & \mathbf{8.0} & \mathbf{4.0} & 1.3          & 1.0          & 3.2 & 0.8 & 5.0           \\
 (A'B')A' &  13.1         & \mathbf{+18.0} & \mathbf{5.9} & 10.7          & 8.8          & 12.9 & 10.9 & 15.1          & 3.2          & \mathbf{8.0} & \mathbf{4.0} & 0.8          & 1.2          & 3.0 & 1.0 & 5.2           \\
  (AB')A' & -12.9         &  \mathbf{18.8} & \mathbf{5.5} & 11.4          & 9.1          & 12.9 & 11.3 & 15.1          & 2.8          & \mathbf{8.7} & \mathbf{4.7} & 1.2          & 1.0          & 2.8 & 1.2 & 5.0
  \end{tabular}
  \end{ruledtabular}
  \label{tab03}
\end{table}

%
%

%